\documentclass[referee]{jfm}

\usepackage{graphicx}
\usepackage{natbib}
\usepackage{subfigure}
\usepackage{amsmath}
\usepackage{amsfonts}
\usepackage{amssymb}
\usepackage{amsbsy} 
\usepackage{amsgen} 
\usepackage{gensymb}
\usepackage{color}
\usepackage{multirow}
\usepackage{wasysym} 

\ifCUPmtlplainloaded \else
  \checkfont{eurm10}
  \iffontfound
    \IfFileExists{upmath.sty}
      {\typeout{^^JFound AMS Euler Roman fonts on the system,
                   using the 'upmath' package.^^J}%
       \usepackage{upmath}}
      {\typeout{^^JFound AMS Euler Roman fonts on the system, but you
                   dont seem to have the}%
       \typeout{'upmath' package installed. JFM.cls can take advantage
                 of these fonts,^^Jif you use 'upmath' package.^^J}%
      }
  \else
  \fi
\fi

\newsavebox{\astrutbox}
\sbox{\astrutbox}{\rule[-5pt]{0pt}{20pt}}

\DeclareFontFamily{U}{euc}{}
\DeclareFontShape{U}{euc}{m}{n}{<-6>eurm5<6-8>eurm7<8->eurm10}{}%
\DeclareSymbolFont{AMSc}{U}{euc}{m}{n} 
\DeclareMathSymbol{\umu}{\mathord}{AMSc}{"16} 

\newcommand{\reynoldsnum}{481}

\title[Experimental study of the influence of anisotropy on turbulence]
{Experimental study of the influence of anisotropy on the inertial scales of turbulence}

\author[K. Chang, G. P. Bewley and E. Bodenschatz]%
{K\ls E\ls L\ls K\ls E\ls N\ns C\ls H\ls A\ls N\ls G,\ns%
G\ls R\ls E\ls G\ls O\ls R\ls Y\ns P.\ns B\ls E\ls W\ls L\ls E\ls Y%
\thanks{Email address for correspondence: gregory.bewley@ds.mpg.de} \break%
\and E\ls B\ls E\ls R\ls H\ls A\ls R\ls D\ns %
B\ls O\ls D\ls E\ls N\ls S\ls C\ls H\ls A\ls T\ls Z%
}

\affiliation{Max Planck Institute for Dynamics and Self-Organization, 
37077 G\"{o}ttingen, Germany\\[\affilskip]
International Collaboration for Turbulence Research}

\pubyear{2010}
\volume{123}
\pagerange{1-20}

\date{26 January 2011 and in revised form \today}

\begin{document}
\maketitle

\begin{abstract}
We ask whether the scaling exponents or the Kolmogorov constants 
depend on the anisotropy of the velocity fluctuations in a turbulent flow with no shear.  
According to our experiment, the answer is no 
for the Eulerian second-order transverse velocity structure function.  
The experiment consisted of 32 loudspeaker-driven jets 
pointed toward the centre of a spherical chamber.  
We generated anisotropy by controlling the strengths of the jets.  
We found that the form of the anisotropy of the velocity fluctuations 
was the same as that in the strength of the jets.  
We then varied the anisotropy, 
as measured by the ratio of axial to radial root-mean-square (RMS) velocity fluctuations, 
between 0.6 and 2.3.  
The Reynolds number was approximately constant at around $R_\lambda$~=~\reynoldsnum.  
In a central volume with a radius of 50~mm, 
the turbulence was approximately homogeneous, axisymmetric, 
and had no shear and no mean flow.  
We observed that the scaling exponent of the structure function 
was $0.70~\pm~0.03$, independent of the anisotropy 
and regardless of the direction in which we measured it.  
The Kolmogorov constant, $C_2$, was also independent of direction and anisotropy 
to within the experimental error of 4\%.  
\end{abstract}


\section{Introduction}
\label{sec:intro}

Theories of turbulence that give precise predictions apply by mathematical necessity 
to statistically isotropic flows.  
A conjecture, due to \cite{kolmogorov:1941a}, 
is that flows characterised by large enough Reynolds 
numbers are locally isotropic, or isotropic at small scales, 
even in the presence of anisotropy at larger scales.  
Understanding the way turbulence tends to local isotropy from various states 
of anisotropy remains an important challenge in the study of turbulence.  
Its importance lies in the fact that almost all flows in the 
natural world and in technical applications are anisotropic, 
and not of sufficiently high Reynolds numbers for its influence to be neglected.  
Any useful theory or model of turbulence should therefore incorporate anisotropy, 
and the purpose of this investigation is to provide test cases for such theories.  

The conjecture of local isotropy has received much attention 
\cite[e.g.][]{saddoughi:1994, kurien:2001, biferale:2005}, 
and has been tested in flows with various kinds of anisotropy.  
Anisotropy can arise in different ways, for example 
through spatial variation of the mean flow \cite*[e.g.][]{tavoularis:1981}, 
or through anisotropy in the velocity fluctuations.  
These different forms of anisotropy might have different effects on the small scales of turbulence.  
No test has been performed previously to isolate the influence of anisotropy in the fluctuations only, 
without also introducing shear.  
Our task was to create a device with which we could do this.  
We show that it is possible in a single apparatus to produce 
turbulence with negligible shear, and with a chosen level of anisotropy on the large scale.  
With the apparatus, we tested the validity of the hypothesis of local isotropy 
under increasingly anisotropic conditions.  

We produced anisotropy in the velocity fluctuations 
by introducing an asymmetry in the agitation of the turbulence.  
Evidence suggests that there is a causal relationship between asymmetry in the agitation of turbulence, 
and anisotropy in the energy-containing scales of turbulence.  
For example, asymmetry and anisotropy are present at the same time in 
turbulent jets \cite*[e.g.][]{hussein:1994}, 
Taylor-Couette flows \cite*[e.g.][]{andereck:1986}, 
and wind tunnels with specially designed shear generators \cite*[e.g.][]{tavoularis:1981}.  
Turbulence produced by computer simulation can also be forced in an asymmetric way; 
\cite*{yeung:1991} studied the influence of this asymmetry.  
One exception is the turbulence produced by a 
grid in a wind tunnel with a contraction, 
where the axis of the tunnel introduces an asymmetry, 
yet the turbulence produced can be nearly isotropic, but decaying \cite[][]{comtebellot:1966}.  
There is also a pattern in the relationship between asymmetry of the forcing 
and anisotropy of the turbulence.  
Machines with a single axis of symmetry, 
such as the von K\'{a}rm\'{a}n flow \cite[e.g.][]{ouellette:2006b} 
and wind tunnels \cite*[e.g.][]{staicu:2003}, 
produced axisymmetric turbulence.  
Machines with more axes, such as the one developed by \cite{hwang:2004}, 
produced isotropic turbulence.  
A careful study by \cite{zimmermann:2010} showed that six axes were 
sufficient to produce turbulence without a preferred direction.  
Our machine had 16 axes.  

We generated both isotropic and axisymmetric flows.  
The apparatus worked by modulating the relative intensity of 32 mixers distributed over a sphere.  
When all the mixers were driven with the same intensity, isotropic turbulence resulted.  
We introduced a preferred axis by driving mixers near one of the 16 
axes of the machine either more strongly or more weakly than the other mixers.  
This produced anisotropy in the turbulence.  
The geometric form of the apparatus was inspired by the one of \cite{zimmermann:2010}, 
except that ours used 32 mixers instead of their 12, 
and air as the working fluid instead of water.  
Our mixers were loudspeakers coupled to nozzles; 
the diaphragm of the loudspeaker pushed and pulled air through an orifice, 
which formed a turbulent jet.  
This technique for mixing air was invented by \cite{hwang:2004}, 
though our implementation differs in several details, described below.  
We used laser Doppler velocimetry (LDV) to measure the fluid velocity 
near the centre of the apparatus.  

We measured the influence of anisotropy on velocity structure functions, 
which are the moments of velocity differences across two points 
separated in space.  The second order structure function is 
\begin{equation}
D_{ij} (\boldsymbol{r})
= \langle (u_i (\boldsymbol{r}) - u_i (0)) (u_j(\boldsymbol{r}) - u_j (0)) \rangle \,, 
\label{eq:D}
\end{equation}
where $u_i (\boldsymbol{r})$ is the fluctuation of the $i$-th 
component of the velocity at location $\boldsymbol{r}$ 
and $\langle \cdot \rangle$ denotes temporal averaging.  
Local isotropy demands that for small enough $\boldsymbol{r}$ 
and large enough Reynolds number, the structure function is isotropic.  
In this case it can be written in terms of a single independent scalar function.  

\cite{kolmogorov:1941a} surmised that in the inertial subrange, 
where he thought that neither the viscosity nor the large-scales 
would influence the dynamics, 
the structure functions follow a power law as a function of spatial separation.  
For representative parts of the structure function tensor given in equation~\ref{eq:D}, 
this can be written 
\begin{align}
D_{zz}(r_1) &\equiv \langle (u_z (r_1) - u_z (0))^2 \rangle 
= \dfrac{4}{3} \, C_2^{(r)} \, (\epsilon \, r_1)^{\zeta_2^{(r)}} \, , 
\label{eq:dzz} \\
D_{r_2r_2}(z) &\equiv \langle (u_{r_2} (z) - u_{r_2} (0))^2 \rangle 
= \dfrac{4}{3} \, C_2^{(z)} \, (\epsilon \, z)^{\zeta_2^{(z)}} \, .  
\label{eq:drr}
\end{align}
These functions are called transverse because 
the velocities, $u_z$ and $u_{r_2}$, are perpendicular to the vectors pointing from 
$0$ to $r_1$ and from $0$ to $z$, respectively.  
We have introduced a certain coordinate system that we describe later.  
Here, $\epsilon$ is the energy dissipation rate per unit mass, 
the dimensionless constants, $C_2^{(r)}$ and $C_2^{(z)}$, 
are called Kolmogorov constants, 
and the scaling exponents, $\zeta_2^{(r)}$ and $\zeta_2^{(z)}$, 
both equal 2/3$^{rds}$ according to \cite{kolmogorov:1941a} for locally isotropic turbulence.  
The Kolmogorov constants are also equal to each other if the turbulence is locally isotropic, 
but their value must be determined by experiment.  
It is interesting to note that the corresponding Kolmogorov constants 
for the Lagrangian structure function vary with angle 
by more than 20\% at a Reynolds numbers of 815  in an anisotropic  
von K\'{a}rm\'{a}n swirling flow \cite[][]{ouellette:2006b}.  
Our work addresses the angular dependence of the Eulerian $C_2$, 
and is in part an extension of the above work.  

One of our measures of anisotropy is the ratio of Kolmogorov constants 
measured in the two different directions, $C_{2}^{(r)} / C_{2}^{(z)}$.  
Variation of this quantity could account for 
differences between published values of the Kolmogorov constant.  
Such differences are evident in \cite{sreenivasan:1995}, 
for example, where values for the constant were collected from various experiments 
and found differ from each other with a standard deviation of about 10\%.  
Could it be that different experiments measured different values for the Kolmogorov constant 
because they had different levels of anisotropy, 
or else because they measured the constant in different directions?  
In our study, we found no dependence of the Kolmogorov constants 
on the anisotropy, as shown in section~\ref{sec:structurefunctions}.  

We also report on the difference between the scaling exponents measured 
in the two directions, $\Delta\zeta_2 = \zeta_2^{(z)} - \zeta_2^{(r)}$.  
Here, we sought to interpret the observation made by \cite{shen:2002} 
that in anisotropic wind tunnel flows 
the scaling exponent in the cross-tunnel direction 
was about 0.1 smaller than the exponent in the direction of the mean flow.  
In our shearless, axisymmetric flow we found, in contrast,  
that the difference between the scaling exponents, $\Delta\zeta_2$, was zero 
to within the measurement accuracy.  
As is explained in section~\ref{sec:structurefunctions}, 
we measured the exponents using the Extended Self-Similarity (ESS) technique 
introduced by \cite{benzi:1993}.  

In the following section~\ref{sec:apparatus}, 
we describe the apparatus and techniques.  
In section~\ref{sec:velocitystats}, 
we show that we could control systematically the anisotropy of the turbulence, 
and report on the extent to which the turbulence had no shear and was axisymmetric.  
A reader with no interest in experimental detail can proceed to section~\ref{sec:structurefunctions}, 
where we report on the measurements of structure functions.  
We conclude in section~\ref{sec:conclusions}.  

\section{The Experiment}
\label{sec:apparatus}

\begin{figure}
\begin{center}
\subfigure[]{
\includegraphics[clip,width=4.7cm]{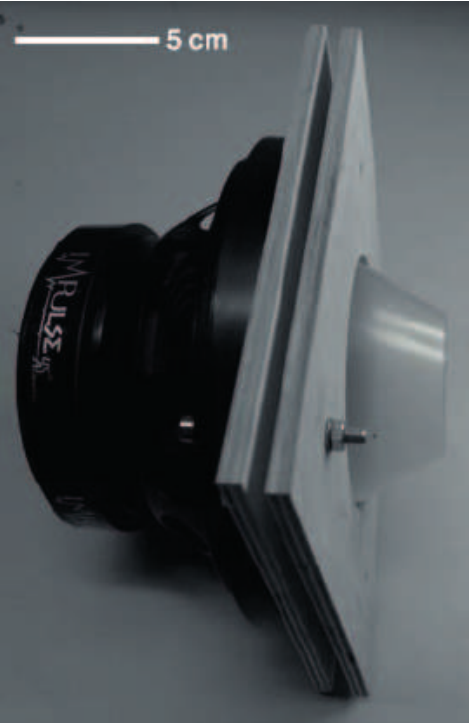}}
\subfigure[]{
\includegraphics[width=8cm]{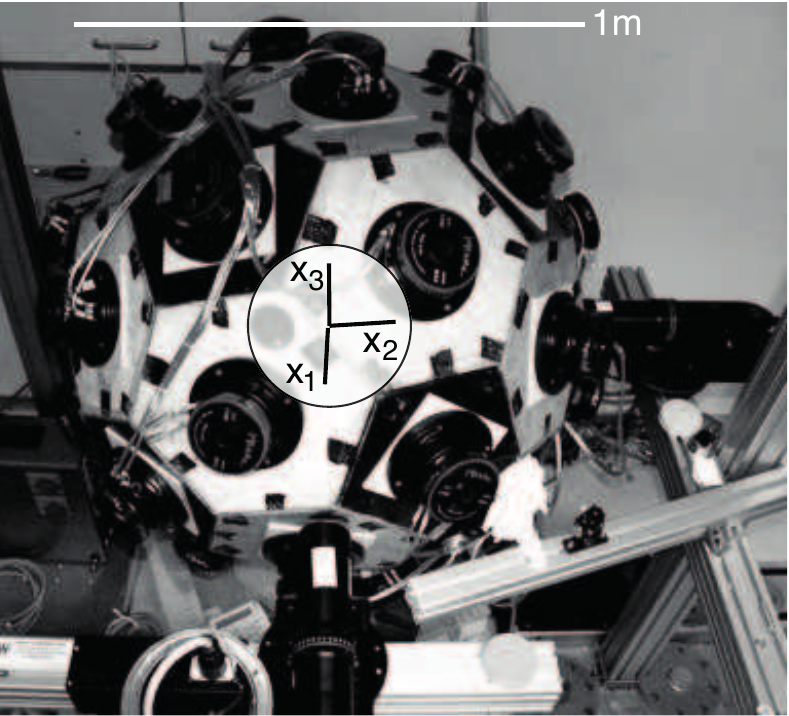}}
\caption{
  (a) The jet generator.  
  (b) The layout of  the turbulence chamber and the LDV probes 
  (black cylinders), one near the bottom of the image, and one to the right.  
  The one aligned with the $x_1$-axis measures 
  the components of particle velocities in both the $x_2$ and $x_3$ directions.  
  The probe aligned with the $x_2$ axis measures velocities in either the 
  $x_1$ or the $x_3$ direction.  
}
\label{fig:soccerball_ldv}
\end{center}
\end{figure}

As shown in figure~\ref{fig:soccerball_ldv} (b), 
the turbulence chamber had the shape of a truncated icosahedron 
and a diameter of 99~cm.  
It was made of wood 

joined together with nylon straps and glue.  
In the centre of each face a circular hole was cut for the jet generator.  
In addition, 
further holes adjacent to the jet generators were cut for optical access.  
As the shape is that of a soccer ball we will use this term 
from now on to describe the turbulence chamber. 

As depicted in figure~\ref{fig:soccerball_ldv}(a), the mixers were jets, 
similar in principle to those developed 
for flow control applications \cite[e.g.][]{glezer:2002}.  
A 150~W loudspeaker with a diameter of 16.5~cm 
and a uniform frequency response between 50 and 3000 Hz 
was mounted on each face of the soccer ball 
and pointed towards its centre.  
The loudspeakers were coupled to 
conical nozzles with opening angle 30$^\circ$, 
length 4.3~cm, and orifice diameter 5~cm.  
When backing away from the orifice the diaphragm drew in air from all directions, 
while it blew it out in a single direction when moving toward the orifice.  
By driving the speaker sinusoidally, we generated a pulsating turbulent jet.  
The sound level inside the soccer ball was typically 135~dB, 
which corresponded to 0.2\% of the measured turbulent kinetic energy.  
Thus we expect the sound to have an negligible effect on the turbulence 
and the measurements of velocity.  

The flow chamber was similar to the ones described by 
\cite*{hwang:2004}, \cite*{webster:2004} and \cite{lu:2008}, who employed 
cubic Plexiglas boxes from which the corners were 
cut off.  
A speaker-driven jet was mounted at each of the eight cut-off corners, 
and pointed toward the centre of the chamber.  
The flow chamber of \cite*{warnaars:2006} was a 
rectangular Plexiglas box with two speakers placed at each square end of the box.  
A grid was placed in front of each speaker to excite small-scale turbulence.  
We compare the flow parameters 
in these apparatuses, as well as the one of \cite{goepfert:2010}, 
in table~\ref{table:flowchambers}.  
\begin{table}
\begin{center}
\def~{\hphantom{0}}
\begin{tabular}{l*{6}l}
& Hwang & Webster & Warnaars & Lu {\it et al.} & Goepfert & Present \\[-0.3cm]
& \& Eaton (2004) & {\it et al.} (2004) & {\it et al.} (2006) & (2008)
& {\it et al.} (2010) & work \\
Geometry & Truncated & Truncated & Rectangular & Truncated & 
Octahedron & Truncated \\[-0.3cm]
& cube & cube & & cube & & icosahedron \\
Flow medium & Air & Saltwater & Water & Air & Air & Air \\
$\sigma$ (ms$^{-1}$) & 0.87 & 0.0097 & 0.00069 & 0.60 & 0.88 & 1.1\\
$\sigma_{u}/\sigma_{v}$ & 1.03 & 1.00 & 0.98 & -- & 0.95 & 0.94 \\
$U / \sigma$ & 0.022 & 0.07 & -- & -- & 0.03 & 0.04 \\
$\epsilon$ (m$^2$s$^{-3}$) & $11$ & $2.5 \times 10^{-5}$ & $ 
1.25 \times 10^{-6}$ & $1.4$ & $5.8$ & $6.7$ \\
$\eta$ ($\umu$m) & 130 & 450 & 950 & 220 & 155 & 155 \\
$\tau_\eta$ (ms) & 1.2 & 200 & 1000 & 3.3 & 1.6 & 1.5 \\
$R_{\lambda}$ & 220 & 68 & 5.9 & 260 & 250 & \reynoldsnum
\end{tabular}
\caption{
  \label{table:flowchambers}
  Chamber geometry and flow statistics from various speaker-driven flows.  
  Symbols from top to bottom: 
  RMS velocity fluctuations ($\sigma$), 
  ratio of RMS velocity fluctuations ($\sigma_{u}/\sigma_{v}$), 
  ratio of mean flow to RMS fluctuations ($U / \sigma$), 
  energy dissipation rate ($\epsilon$), 
  Kolmogorov length scale ($\eta$), 
  Kolmogorov time scale ($\tau_\eta$), Taylor Reynolds number ($R_\lambda$).  
  For each study, we give the highest Reynolds number, lowest mean flow and 
  best isotropy reported by the authors.  
  For our apparatus, we report only on data acquired under isotropic forcing.  
}
\end{center}
\end{table}

\begin{figure}
\begin{center}
\includegraphics[clip,width=3.2in]{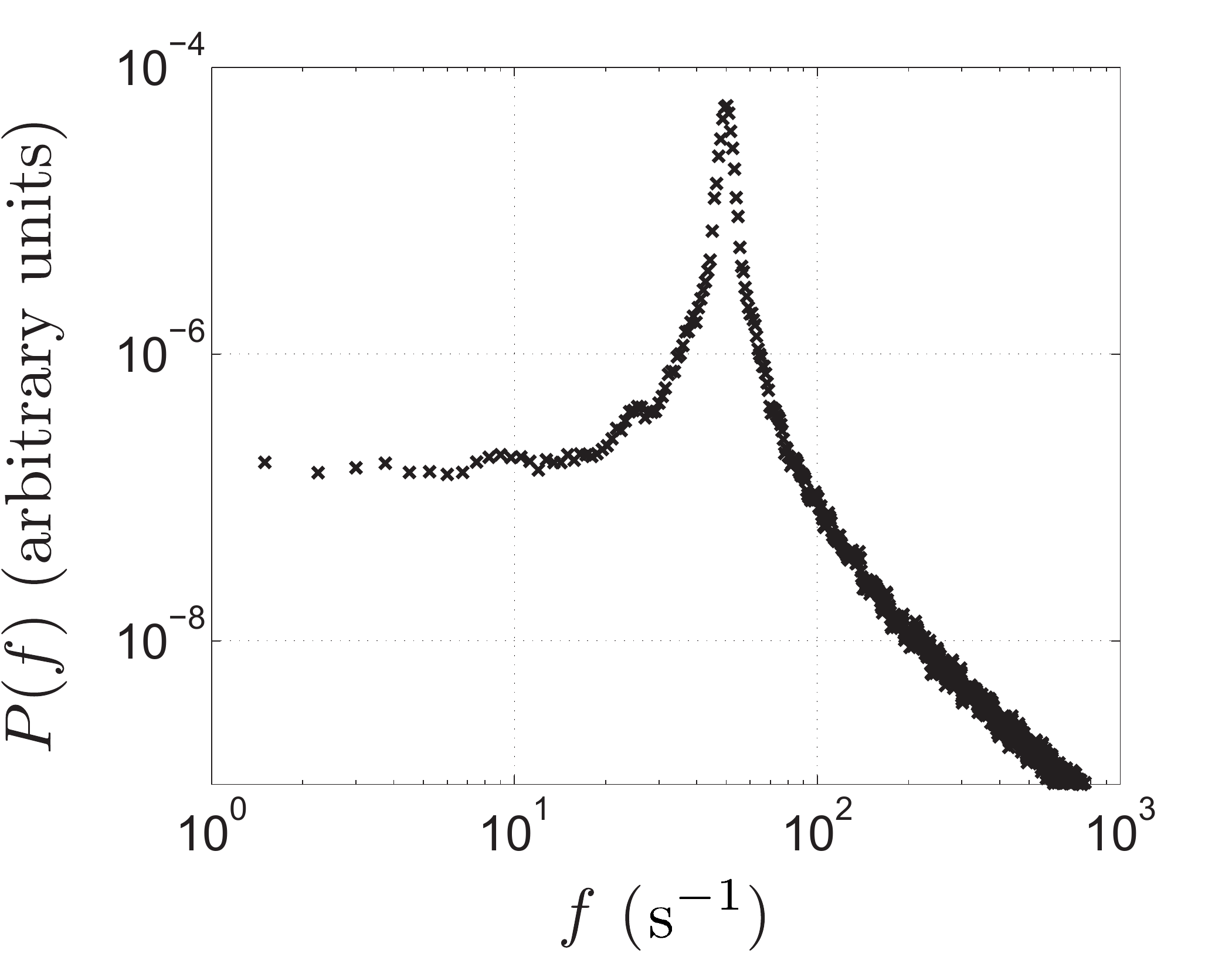}
\caption{
The power spectrum of the voltage applied to the loudspeakers.  
}
\label{fig:stirringspectrum}
\end{center}
\end{figure}

We chose the signals that drove the loudspeakers 
to generate the desired anisotropy and low mean velocity in the flow.  
Figure~\ref{fig:stirringspectrum} shows the power spectrum of the signal.  
The 50~Hz amplitude-modulated sine signals 
were digitally generated at a sampling frequency of  3~kHz.  
All speakers were driven in phase, and 50~Hz produced the strongest jets.  
The 50~Hz oscillation was not, however, detectable in the spectrum of the fluid velocities 
measured near the middle of the soccer ball.  
We modulated the amplitude of each driving signal with noise \cite[][]{fox:1988}, 
though other forms of noise would probably also work.  
The correlation time was 0.1 seconds, 
which was  approximately equal to the large-scale eddy turn-over time, 
$L/\sigma$, $L$ being a characteristic length scale describing the 
large-scale motions of the flow, and $\sigma$ being the root mean squared 
(RMS) velocity fluctuations.  
This condition ensured that fluctuations in the energy input rate 
to the turbulence occurred on time scales equal to or faster 
than the turbulence decay time, so that the turbulence was in a steady state.  
In any case, we found that the statistical properties of the 
turbulence were insensitive to the correlation time.  

To avoid mean flows, the amplitudes of each driving signal were 
adjusted so that the sum of the amplitudes of all signals was zero.  
This also reduced the amplitude of the sound generated 
and the air exchange between the inside of the 
ball and the room.  

We selected the desired anisotropy by separately 
adjusting the RMS amplitude of each speaker 
As we restricted ourselves to cylindrically symmetric driving, 
the asymmetry could be characterised by the ratio of the axial 
amplitude, $a_{\rm axial}$, to the radial amplitude, $a_{\rm radial}$: 
\begin{equation}
A = \dfrac{a_{\rm axial}}{a_{\rm radial}} \,.  
\label{eq:aspectratio}
\end{equation}
A forcing is then described as exhibiting oblate spheroidal asymmetry when 
$0 < A < 1$, spherical symmetry when $A=1$ 
and prolate spheroidal asymmetry when $A>1$. 

As we have 32 speakers distributed with icosahedral symmetry, 
the RMS amplitude for a given loudspeaker was set according to the following scheme.  
We defined an ellipsoid that had the same centre as the soccer ball, 
and whose major axis to minor axis ratio was $A$.  
We then calculated the intersection between the surface of the ellipsoid 
and the vector that pointed from the centre of the soccer ball 
to the centre of the given loudspeaker.  
The distance between this intersection and the centre of the soccer 
ball set the relative RMS amplitude for the given loudspeaker.  

We characterised the flow using a TSI laser Doppler velocimetry (LDV) system.  
We used oil droplets as tracer particles, generated by a Palas AGF 10.0 aerosol generator.  
The droplets had a most probable diameter of about 3~$\umu$m, 
and a volume fraction of less than $10^{-7}$.  
Droplets of this diameter settle in still air at 200~$\umu$m~s$^{-1}$, 
which was much smaller than their RMS velocities in the turbulent flow.  
These oil droplets were 
passive tracers, since the Stokes number in our flow was very small 
(approximately 0.02) \cite*[e.g.][]{bewley:2008} 
and the accelerations of the turbulent flow were much larger than 
the acceleration of gravity \cite[e.g.][]{voth:2002}.  

As shown in figure~\ref{fig:soccerball_ldv}(b), we used two LDV probes.  
One probe measured two components, 
and the other only one component of the velocity of individual tracer particles.  
The measurement volume was approximately ellipsoidal in shape, 
and was approximately 100~$\umu$m in diameter and 2~mm in length, 
which allowed us to 
resolve scales larger than the dissipation scale of the turbulence.  
The mean data sampling rate was between 300 and 3000 samples per second, 
depending on the probe.
We typically collected $2\times 10^{6}$ data points per aspect ratio 
of the forcing and per spatial position of the probes.  

LDV data are known to suffer from biases, since the system observes 
particles with larger velocities more often than those with smaller 
velocities \cite[see e.g.][]{albrecht:2003}.  
We used the residence time weighting 
suggested by \cite{buchhave:1975} and \cite*{buchhave:1979} 
to correct our single point statistics.  
We did not correct for possible bias in our two-point statistics.  
To calculate the two-point statistics the velocity signals were first 
processed using a slotting technique, i.e., we searched for samples coming 
from each of the two probes with time separations falling within a 
temporal bin of 1~ms duration, and 
within each temporal bin, the data from each probe was averaged.  
Approximately $10^4 - 10^5$ points contributed to the statistics for each spatial separation.  
We found that as long as the temporal bin was smaller than 15~ms 
the difference in the results was less than 1\%.  

\begin{figure}
\begin{center}
\includegraphics[clip,width=2.5in]{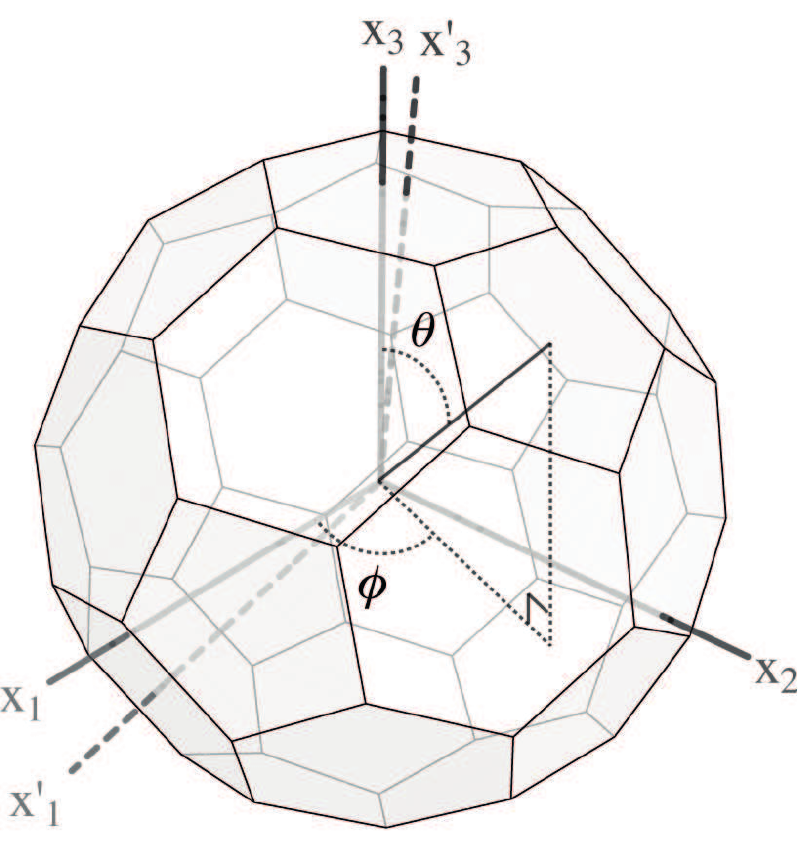}
\caption{
  The schematic shows the coordinate system in the laboratory frame.  
  $\hat{x}_3$ was aligned with the vertical.  
  The axis of symmetry of the forcing lay along 
  either $x^{\prime}_1$ or $x^{\prime}_3$, depending on the experiment.  
  Our measurements were made at points along $x_1$.  
}
\label{fig:labcoordinates}
\end{center}
\end{figure}

Before we proceed to the measurements, we describe the coordinate system, 
which was fixed to the axis of symmetry of the turbulence.  
Please note that the coordinate 
system was not fixed in the laboratory frame, as is shown in 
figures~\ref{fig:soccerball_ldv} and \ref{fig:labcoordinates}.  
Axisymmetric turbulence has two principal axes.  
However, our measurement apparatus measured velocities only at 
points that lay along a single line in the laboratory frame.  
In order to sample the statistics of  turbulence along both axes, we rotated the 
axis of symmetry of the turbulence by taking advantage of the symmetries of the soccer ball. 

The soccer ball was oriented such that one of the symmetry axes of the forcing 
lay along $\theta = 107^{\circ}$ 
and $\phi = -4^{\circ}$, which is labelled as $x^{\prime}_1$ in 
figure~\ref{fig:labcoordinates}, and is close to the $x_1$-axis.  
The other symmetry axis lay along $\theta = 4^{\circ}$ and 
$\phi = 180^{\circ}$, which is labelled as $x^{\prime}_3$ in 
figure~\ref{fig:labcoordinates}, and is close to the $x_3$-axis.  
Because the primed and unprimed coordinate systems are close to each other, 
we do not distinguish between them in the rest of this paper.  
This simplification does not impact our conclusions, 
since we find no difference in the small-scale statistics between the two directions, 
despite the fact that they approximately orthogonal.  

\begin{figure}
\begin{center}
\subfigure[]{\includegraphics[width=6.5cm]{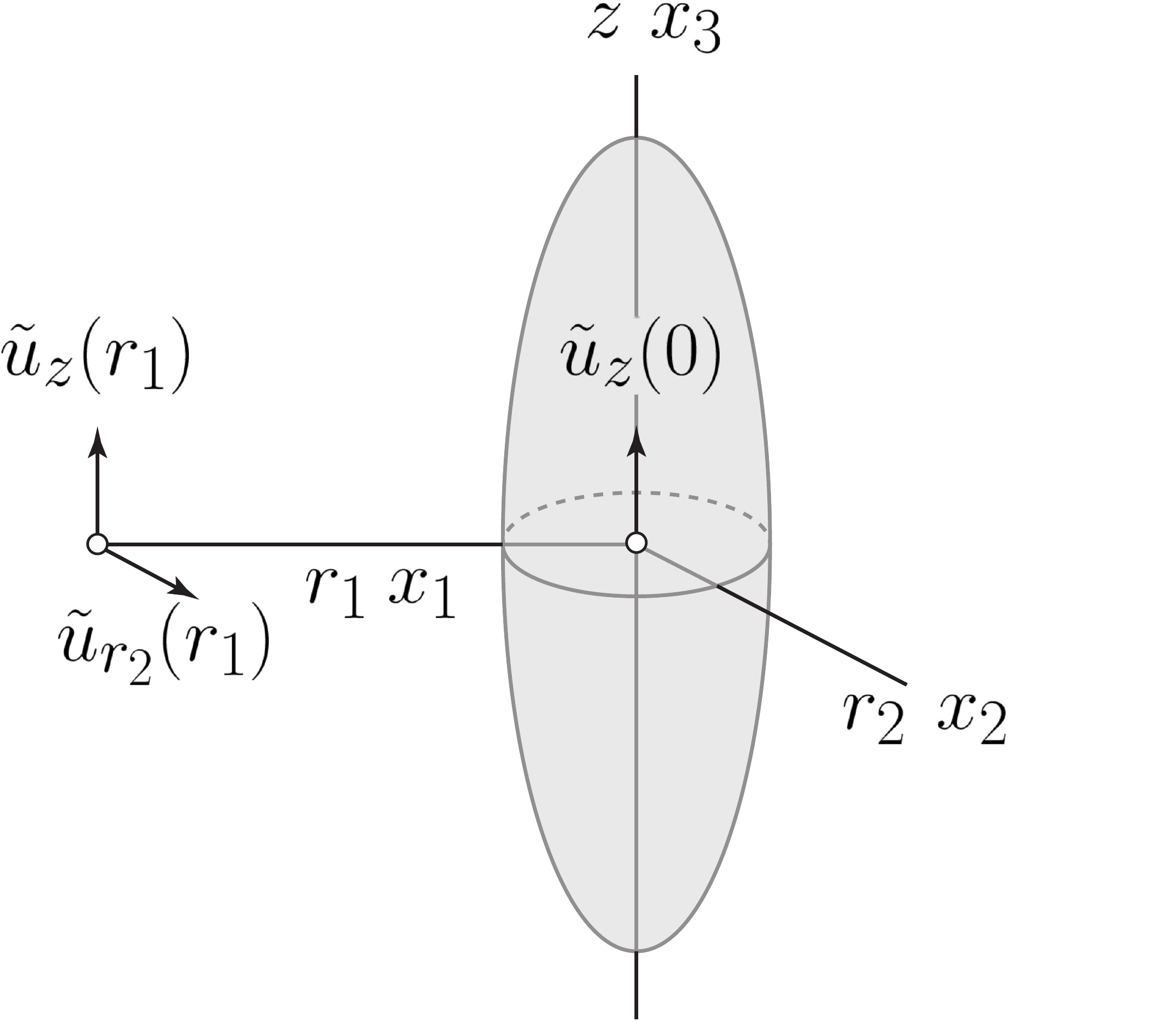}}
\subfigure[]{\includegraphics[width=6.5cm]{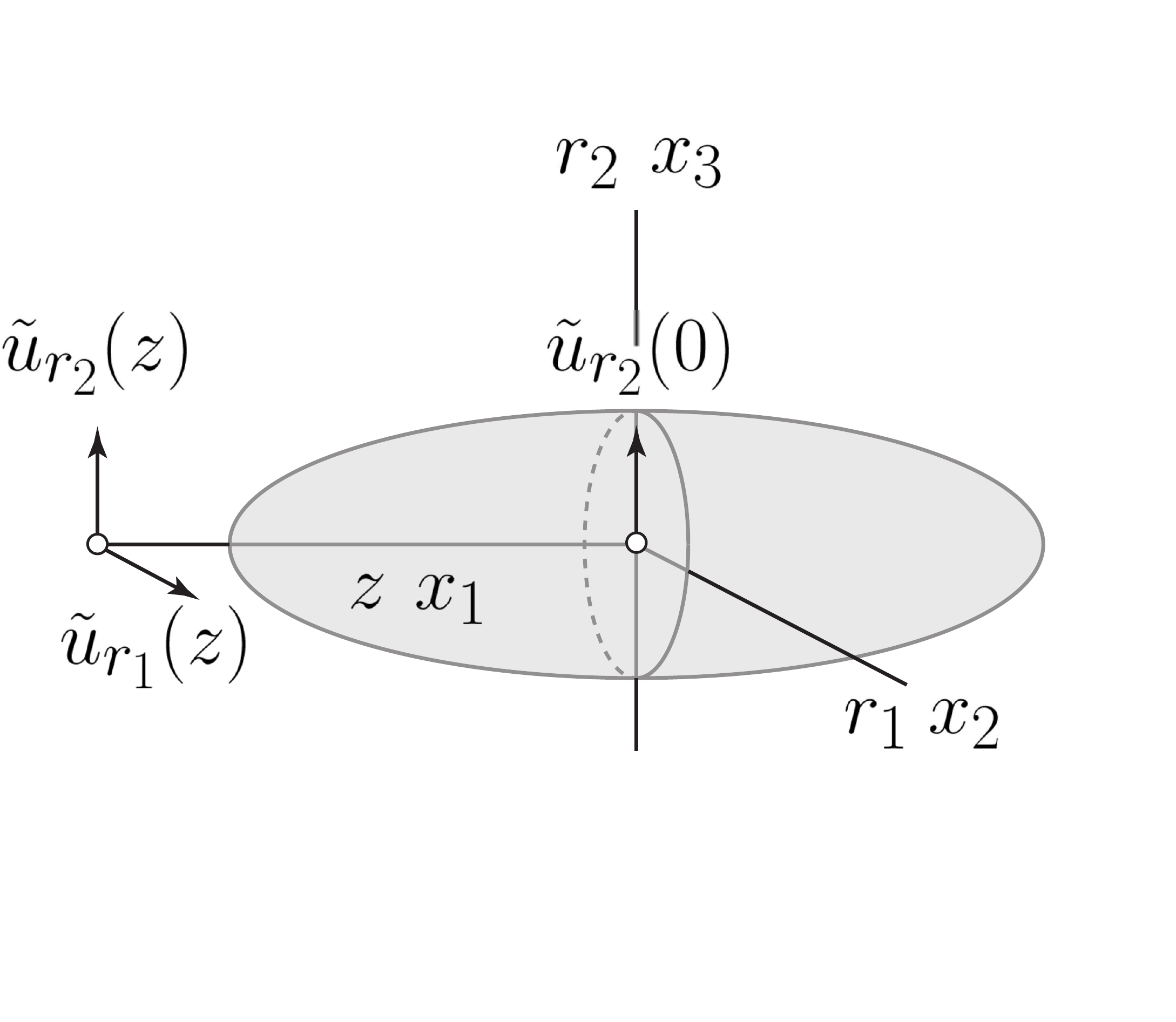}}
\caption{
  Orientations of the body coordinate system of the 
  forcing with respect to the laboratory frame, and the 
  conventions we use (a) when the axis of symmetry of the forcing lay 
  close to the $x_3$ axis of the laboratory frame, and (b) when the
  axis of symmetry lay close to $x_1$ axis of the laboratory
  frame.  The coordinate system $(x_1, x_2, x_3)$ 
  is fixed in  the laboratory frame, and the
  coordinate system $(r_1, r_2, z)$ is fixed with respect to the
  symmetry of the forcing.  
}
\label{fig:bodycoordinates}
\end{center}
\end{figure}

Figure~\ref{fig:bodycoordinates} shows the coordinate system, 
$(r_1, r_2, z)$, which was aligned with the symmetry axis of the forcing.  
The coordinate system has two orientations with respect to the laboratory 
coordinate system, corresponding to the two cases described above.  
Hereafter, `axial' refers to both the direction of the velocity 
component measured along the axis of symmetry, and, in discussing 
two-point statistics, separations along the axis of symmetry.  
Similarly, `radial' refers to both the direction of the velocity 
components perpendicular to the axis of symmetry, and to separations 
perpendicular to the axis of symmetry.  In addition, we refer to 
two-point quantities whose separation vector lies along the axis of 
symmetry as axial quantities, and those with radial separations 
as radial quantities.  For example, $f (x_1)$ 
is denoted as $f(z)$ when the axis of symmetry of the 
forcing lay along $x_1$, and is called an axial quantity.  
In keeping with the geometry described above, our `axial' measurements were in reality about 15$^\circ$ away from the axis of symmetry, and the `radial'  measurements were about 4$^\circ$ from its normal.  

In the first part of this paper, we discuss measurements made at a single point 
of all three orthogonal components  of velocity.  
For these measurements, 
both probes observed a fixed volume (aligned to within 10~$\umu$m) 
close to the centre of the soccer ball, 
which is the point $(0,0,0)$ in the coordinate system shown in 
figure~\ref{fig:labcoordinates}.  
One probe measured a 
single component of the particle velocities, namely $\tilde{u}_{x_1} (0,0,0)$.  
The second probe measured two 
orthogonal components of the particle velocities, namely $\tilde{u}_{x_2} (0, 0, 0)$ 
and $\tilde{u}_{x_3} (0, 0, 0)$.  
For measurements described in the second part of this paper, 
we aligned the probes in a similar way, except that we positioned the 
second probe at different stations along the $x_1$ axis, using a 
programmable linear traverse.  This probe now measured 
$\tilde{u}_{x_2} (x_1, 0, 0)$ and $\tilde{u}_{x_3} (x_1, 0, 0)$.  
In addition, the single-velocity-component probe was rotated $90\degree$ 
to measure $\tilde{u}_{x_3}(0,0,0)$, which was coincident with one of the 
components measured by the two component probe when $x_1$ 
equaled zero.  

For each of a series of forcing anisotropies, $A$, given by 
equation~\ref{eq:aspectratio}, we collected two data sets.  
For one data set, we aligned the axis of symmetry of the forcing 
close to $x_3$, and for the other, we aligned the symmetry axis close to $x_1$.  
For each value of the large-scale anisotropy, and each orientation of the 
symmetry axis, we collected data with the two-component LDV probe 
stationed at various positions along the $x_1$-axis.  
We varied the anisotropy while fixing the quantity 
\begin{equation}
K = \frac{1}{2} \, \Big[\langle u_{x_1}^{2} (0,0,0) \rangle + 
\langle u_{x_2}^{2} (0,0,0) \rangle + \langle u_{x_3}^{2}
(0,0,0) \rangle \Big] \, , 
\label{eq:K}
\end{equation}
where the fluctuating parts of the velocities, $\tilde{u}_i$, are denoted by $u_i$, 
and $\langle \cdot \rangle$ denotes temporal averaging.  
In other words, we fixed the turbulence kinetic energy in the centre of the ball.  

\section{Anisotropy, axisymmetry and homogeneity}
\label{sec:velocitystats}

Three-dimensional velocity measurements made at 
single points in the centre region of the soccer ball show that the flow was 
approximately homogeneous and axisymmetric.  
Furthermore, 
the anisotropy of the turbulent velocity RMS fluctuations followed the anisotropy of the forcing signal.  
For the case of spherically symmetric forcing, 
with $A$~=~1 (see equation~\ref{eq:aspectratio}), 
we expected and indeed found that 
the turbulence was isotropic.  
In this case, the components of the fluctuating velocity 
were each approximately 1~m~s$^{-1}$,  
and the mean velocity was less than 0.15~m~s$^{-1}$, or 
less than 15\% of the fluctuations, as shown in 
Table~\ref{table:velocityisotropic}.  
\begin{table}
\begin{center}
\begin{tabular}{crr}
\multirow{2}{*}{Velocity Statistics (m~s$^{-1}$)} & \multicolumn{2}{c}{Axis
  Orientation} \\
& $z^{\prime}$ & $x^{\prime}$ \\
$U_{x_1} (0,0,0)$ & $0.099$ & $0.14$ \\
$U_{x_2} (0,0,0)$ & $0.0055$ & $-0.0044$ \\
$U_{x_3}(0,0,0)$ & $-0.038$ & $0.027$ \\
$\sigma_{x_1} (0,0,0)$ & $0.99$ & $0.99$ \\
$\sigma_{x_2} (0,0,0)$ & $0.98$ & $0.98$ \\
$\sigma_{x_3} (0,0,0)$ & $0.96$ & $0.98$
\end{tabular}
\caption{
  \label{table:velocityisotropic}
  Three-component velocity statistics for symmetric forcing, 
  measured at the centre of the soccer ball.  
  The symbols $U$ indicate the mean flow and the symbols $\sigma$ 
  indicate the RMS velocity fluctuations.  
}
\end{center}
\end{table}

\begin{figure}
\begin{center}
\includegraphics[scale=1]{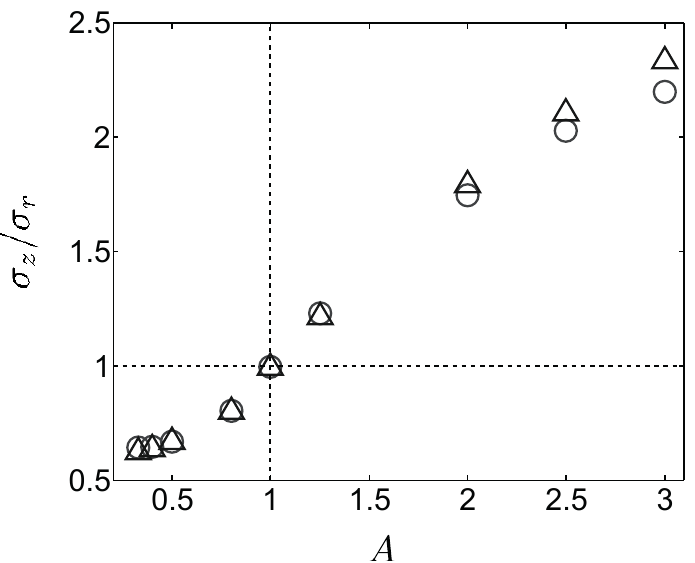}
\caption{
    Velocity fluctuation ratios as a function of the 
    forcing anisotropy, $A$.  $\ocircle$ is $\sigma_{z} / 
    \sigma_{r_1}$ and $\bigtriangleup$ is $\sigma_{z}
    / \sigma_{r_2}$.  Data were 
    collected with the axis of symmetry lying along $x_{3}^{\prime}$.  
    We obtained similar results (not shown) when the axis of symmetry 
    lay along $x_{1}^{\prime}$.  
}
\label{fig:uratio_aspectratio}
\end{center}
\end{figure}

Figure~\ref{fig:uratio_aspectratio} shows the ratio 
of axial to radial velocity  fluctuations as a function of the 
forcing anisotropy, $A$.  
Hereafter, we refer to $\langle u_{i}^{2} (z,r_1,r_2) \rangle^{1/2}$ as 
$\sigma_i (z,r_1,r_2)$, and $\sigma_i$ as $\sigma_i (0,0,0)$.  
Clearly the  anisotropy of the turbulence followed the 
variation of $A$.  
In addition, the anisotropy was nearly the same whether measured
in the $r_1$ or $r_2$ directions.  
Thus the turbulence was close to cylindrically symmetric, 
though this was less so at extreme values of $A$.  
The degradation in the cylindrical symmetry might be explained by 
the fact that the number 
of loudspeakers doing most of the work to drive the turbulence decreased 
as the value of $A$ moved away from one.  
This is the nature of the forcing algorithm described before.  
As the number of loudspeakers effectively decreased, the turbulence probably 
became more sensitive to mechanical 
differences between the speakers, and to misalignments of the nozzles.  

\begin{table}
\begin{center}
\begin{tabular}{l*{8}{r}}
$A$ & 0.33 & 0.4 & 0.5 & 0.8 & 1.0 & 1.25 & 2.0 & 2.5 \\
$\sigma_z$ (m~s$^{-1}$) & 
0.74 & 0.75 & 0.79 & 0.91 & 1.08 & 1.21 & 1.42 & 1.47 \\ 
$\sigma_{r_2}$ (m~s$^{-1}$) & 
1.26 & 1.24 & 1.27 & 1.19 & 1.15 & 1.03 & 0.83 & 0.75 \\
$\sigma_{z} / \sigma_{r_2}$ & 
0.59 & 0.60 & 0.63 & 0.77 & 0.94 & 1.17 & 1.71 & 1.97 \\
$K$ (m$^2$~s$^{-2}$) & 
1.87 & 1.83 & 1.92 & 1.82 & 1.91 & 1.79 & 1.69 & 1.65 \\ 
$|U_{z} / U_{r_2}|$ & 
0.31 & 0.16 & 0.13 & 0.15 & 4.04 & 2.57 & 7.58 & 23.9 \\
$|U|$ (m~s$^{-1}$) & 
0.17 & 0.18 & 0.13 & 0.10 & 0.05 & 0.07 & 0.14 & 0.24 \\
$\langle u_{r_1} \, u_{r_2} \rangle / (\sigma_{r_1} \, \sigma_{r_2})$ & 
$-0.07$ & $-0.06$ & $-0.06$ & $-0.07$ & $-0.07$ & $-0.05$ & $0.01$ & $0.04$ \\
$\langle u_{z} \, u_{r_2} \rangle / (\sigma_{z} \, \sigma_{r_2})$ & 
$-0.06$ & $-0.07$ & $-0.06$ & $-0.07$ & $-0.07$ & $-0.06$ & $-0.01$ & $0.02$ \\
$\epsilon$ (m$^2$~s$^{-3}$) & 
5.4 & 4.9 & 5.4 & 6.0 & 6.7 & 7.0 & 6.5 & 6.4 \\ 
$\eta$ ($\umu$m) & 
163 & 167 & 163 & 159 & 155 & 153 & 156 & 156 \\
$\tau_{\eta}$ (ms) & 
1.7 & 1.8 & 1.7 & 1.6 & 1.5 & 1.5 & 1.6 & 1.6
\end{tabular}
\caption{
  \label{table:velstat}
  The table shows the turbulence 
  parameters, for various loudspeaker RMS amplitude ratios ($A$).  
  The quantity $K$ is the turbulent kinetic 
  energy, $\frac{1}{2}(\sigma^{2}_z + 2 \, \sigma^{2}_{r_2})$, 
  and $|U|$ is $(U_z^2 + 2 \, U_{r_2}^2)^{1/2}$.  
  Data were collected at the centre of the soccer ball 
  when the axis of symmetry was horizontal.  
  We obtained similar results (not shown) when the axis of symmetry was vertical.  
}
\end{center}
\end{table}

\begin{figure}
\begin{center}
\subfigure[]{
\includegraphics[scale=1]{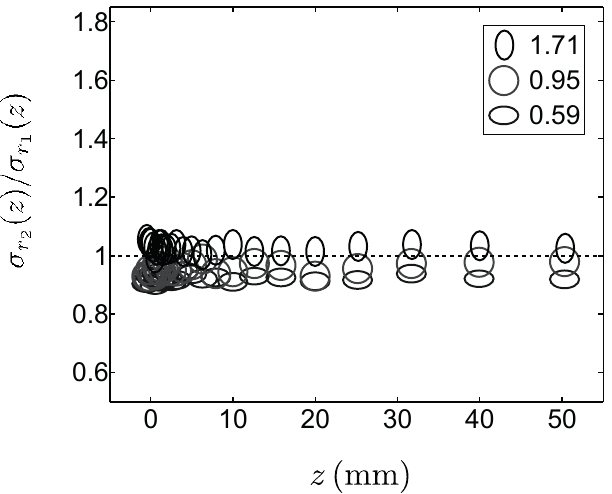}}
\hfill
\subfigure[]{
\includegraphics[scale=1]{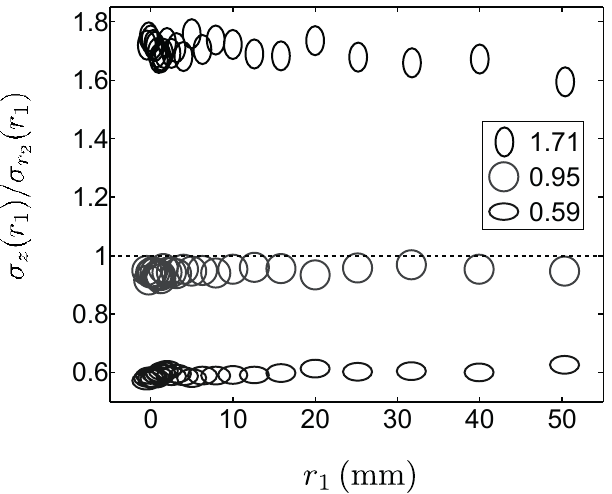}}
\caption{
 To demonstrate homogeneity of the turbulence axisymmetry, 
 we show (a) the ratio of the two radial components of the velocity fluctuations, 
 $\sigma_{r_2} (0,0,z) / \sigma_{r_1} (0,0,z)$, 
 and (b) the ratio of the axial fluctuations to the radial fluctuations, 
 $\sigma_{z} (r_1,0,0) / \sigma_{r_2} (r_1,0,0)$, 
 at various stations moving away from the centre of the ball.  
 The values in the legends, and in all other legends in this paper, 
 are those of the data in (b) for $r_1 = 0$.  
}
\label{fig:vel_anisotropy}
\end{center}
\end{figure}

\sloppy

Next, we consider the degree to which the axisymmetry was spatially uniform.  
To test this, we evaluate how the velocity fluctuation ratios vary with distance 
from the centre of the soccer ball.  
Figure~\ref{fig:vel_anisotropy}(a) shows the ratio of the two radial fluctuating velocities, 
$\sigma_{r_2} (0,0,z) / \sigma_{r_1} (0,0,z)$, 
at various distances from the centre of the ball along the axial direction.  
Within 50~mm, the values of this ratio deviated by less than 10\% from 1, 
which indicates that the turbulence was close to cylindrically symmetric.  
Figure~\ref{fig:vel_anisotropy}(b) shows 
the ratio of the axial fluctuating velocity to one of the radial fluctuating velocities, 
$\sigma_{z} (r_1,0,0) /  \sigma_{r_2} (r_1,0,0)$.  
The values at $r_1 = 0$ correspond to the values shown 
in figure~\ref{fig:uratio_aspectratio}.  
Again, the velocity fluctuation ratio is approximately constant 
within 50~mm of the centre of the ball.  
We infer that the anisotropy was approximately 
uniform within a central region with radius 50~mm.  
\begin{figure}
\begin{center}
\subfigure[]{
\includegraphics[scale=1]{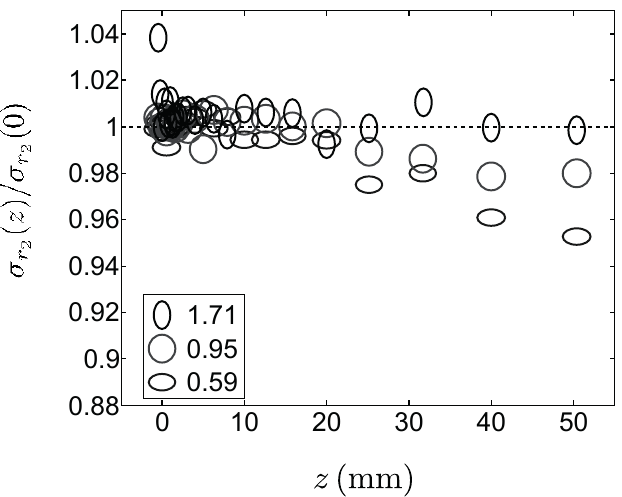}}
\hfill
\subfigure[]{
\includegraphics[scale=1]{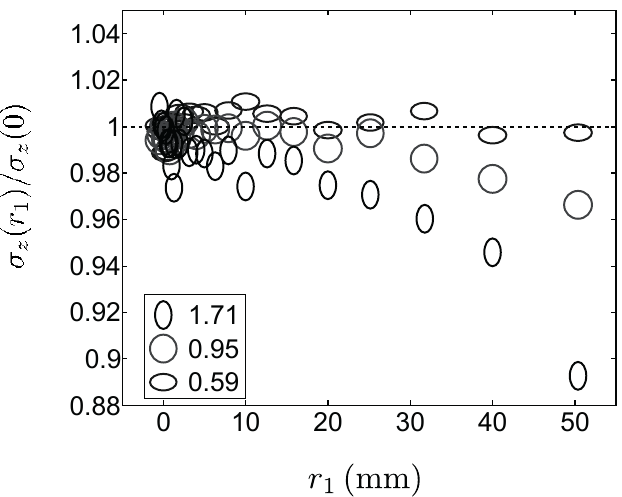}}
\caption{
To demonstrate homogeneity of the turbulent fluctuations, 
we show (a) the radial fluctuations at positions along the axis, 
$\sigma_{r_2} (0,0,z) / \sigma_{r_2} (0,0,0)$, 
and (b) the axial fluctuations at positions normal to the axis, 
$\sigma_{z} (r_1,0,0) / \sigma_{z} (0,0,0)$.  
As in figure~\ref{fig:vel_anisotropy}, the values in the legend 
are the anisotropies measured at the centre of the soccer ball.  
}
\label{fig:vel_homogeneity}
\end{center}
\end{figure}

To evaluate the homogeneity of the turbulent fluctuations, 
we show measurements of 
$\sigma_{r_2} (0,0,z) / 
\sigma_{r_2} (0,0,0)$ and 
$\sigma_z (r_1,0,0) / 
\sigma_{z} (0,0,0)$ 
in figures~\ref{fig:vel_homogeneity}(a) and (b).  
These quantities compare the amplitudes of the velocity fluctuations at locations 
away from the centre to those in the centre.  
Within a radius of 50~mm, the difference between 
$\sigma_{r_2} (0,0,z) / \sigma_{r_2} (0,0,0)$ 
and its value at the origin was within 5\% of the value at the origin.  
The inequality also held for 
$\sigma_{z} (r_1,0,0) / \sigma_{z} (0,0,0)$.  

\fussy

\begin{figure}
\begin{center}
\subfigure[]{
\includegraphics[width=6.25cm]{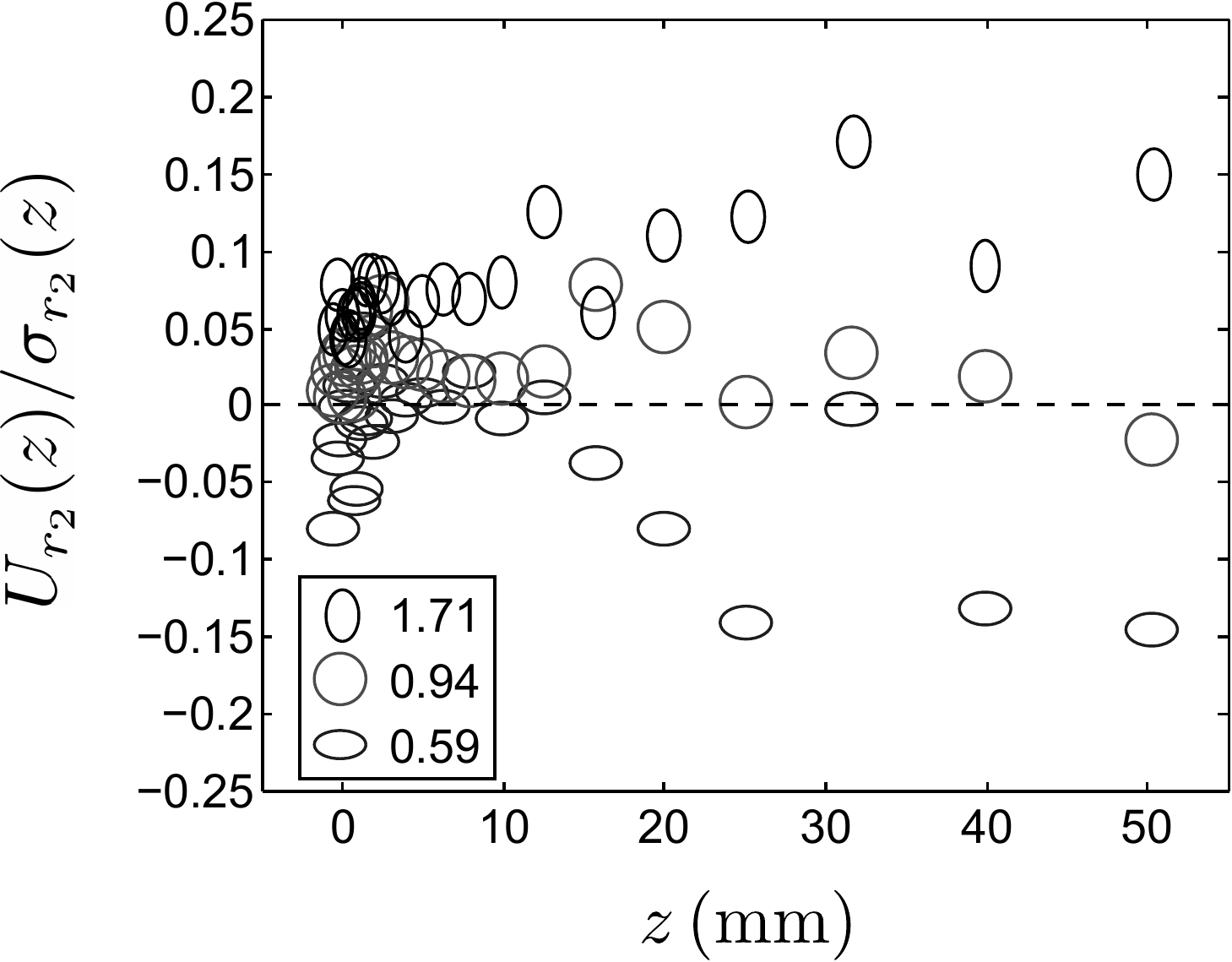}}
\hfill
\subfigure[]{
\includegraphics[width=6.25cm]{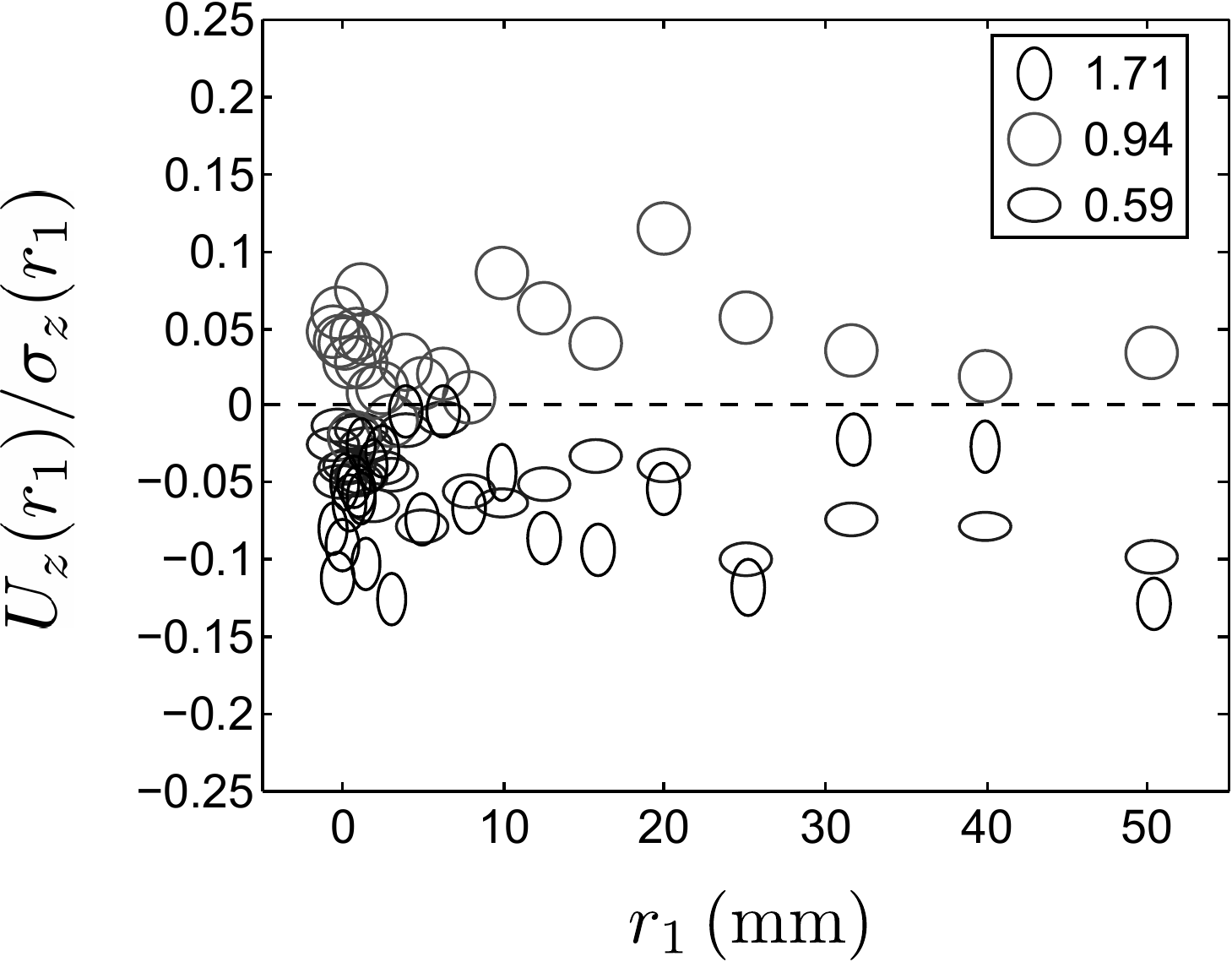}}
\caption{
  The mean flow in two directions as a fraction of the fluctuating velocities in those directions.  
  The values in the legend are the anisotropies measured at the centre of the soccer ball.  
}
\label{fig:mean_to_fluc_homogeneity}
\end{center}
\end{figure}

Anisotropy can manifest itself not only in the fluctuations, 
but also through spatial variation of the mean velocity.  
In figures~\ref{fig:mean_to_fluc_homogeneity}(a) and (b),  
we show the ratios of the mean flow velocities to the velocity fluctuations, 
$U_{r_2} (0,0,z) / \sigma_{r_2} (0,0,z)$ 
and 
$U_{z}(r_1,0,0) / \sigma_{z} (r_1,0,0)$.  
The shape of the curves are difficult 
to interpret, but may indicate that the natural centre of the turbulence 
was offset from the centre of our coordinate system.  
We concluded that variation in the mean velocity was negligible 
because it was typically more than ten times smaller in magnitude than the fluctuations, 
so that the energy in the mean flow was less than 1\% of that in the fluctuations.  
Only at the extreme values of 
the anisotropy was the mean velocity as much as 15\% of the 
fluctuations, which may have been due to the sensitivity of the 
turbulence to small differences between the speakers in this range 
of anisotropies, as discussed above.  

Turbulence production by a mean shear is gauged by the Reynolds stresses, 
or the cross correlations between orthogonal velocity components.  
We measured two such stresses, 
$\langle u_{r_1} (0,0,z) \, u_{r_2} (0,0,z) \rangle$ 
and $\langle u_{z} (r_1,0,0) \, u_{r_2} (r_1,0,0) \rangle$, 
at various locations, $z$ and $r_1$, along the two axes.  
The stresses were less than 7\% of the corresponding kinetic energies, 
$\sigma_{r_1} \, \sigma_{r_2}$ and 
$\sigma_{z} \, \sigma_{r_2}$, 
respectively, and so had a only a small influence on the flow.

\section{Universality in the second order velocity structure functions}
\label{sec:structurefunctions}

\begin{figure}
\begin{center}
\subfigure[]{
\includegraphics[scale=0.6]{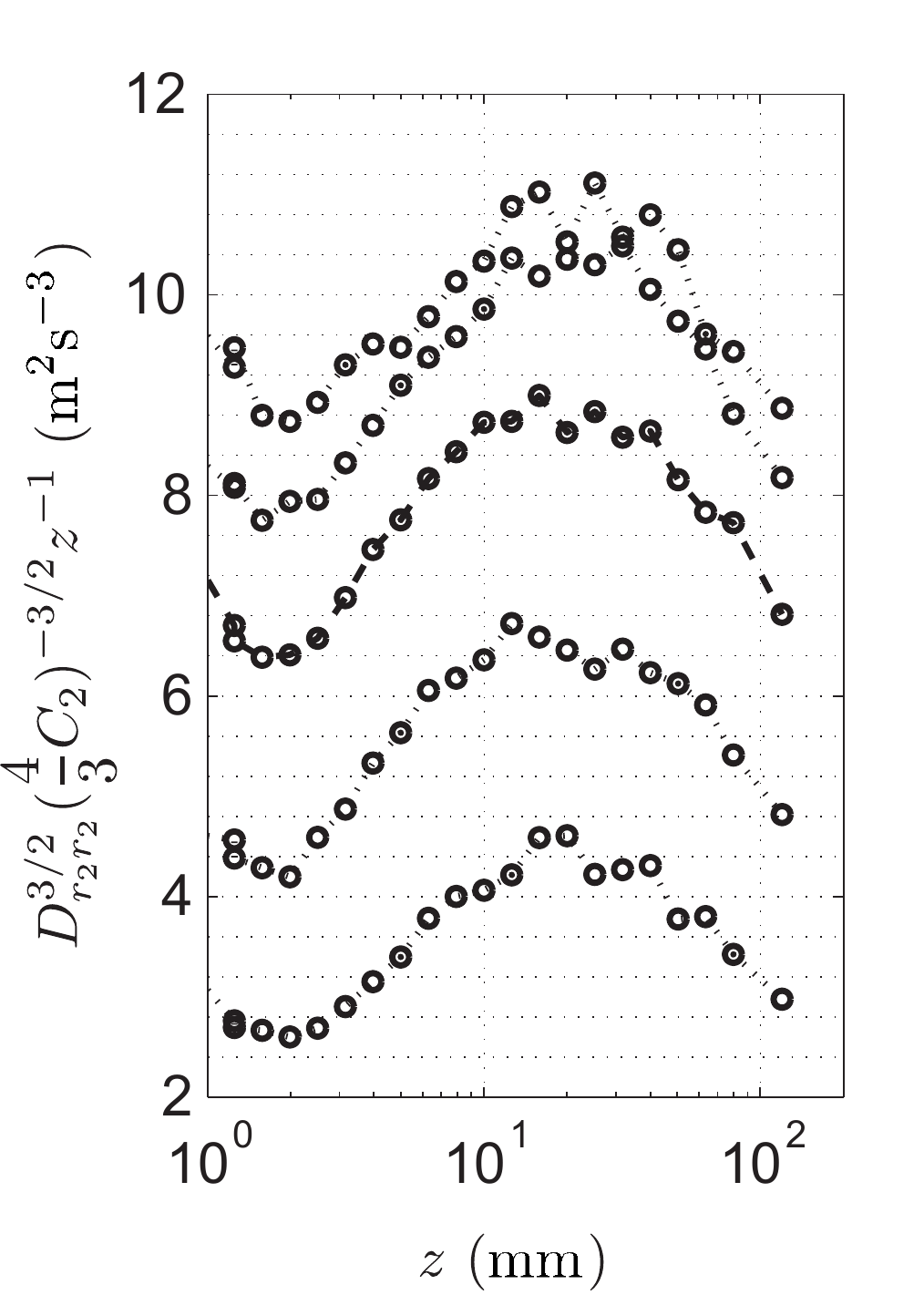}}
\hfill
\subfigure[]{
\includegraphics[scale=0.6]{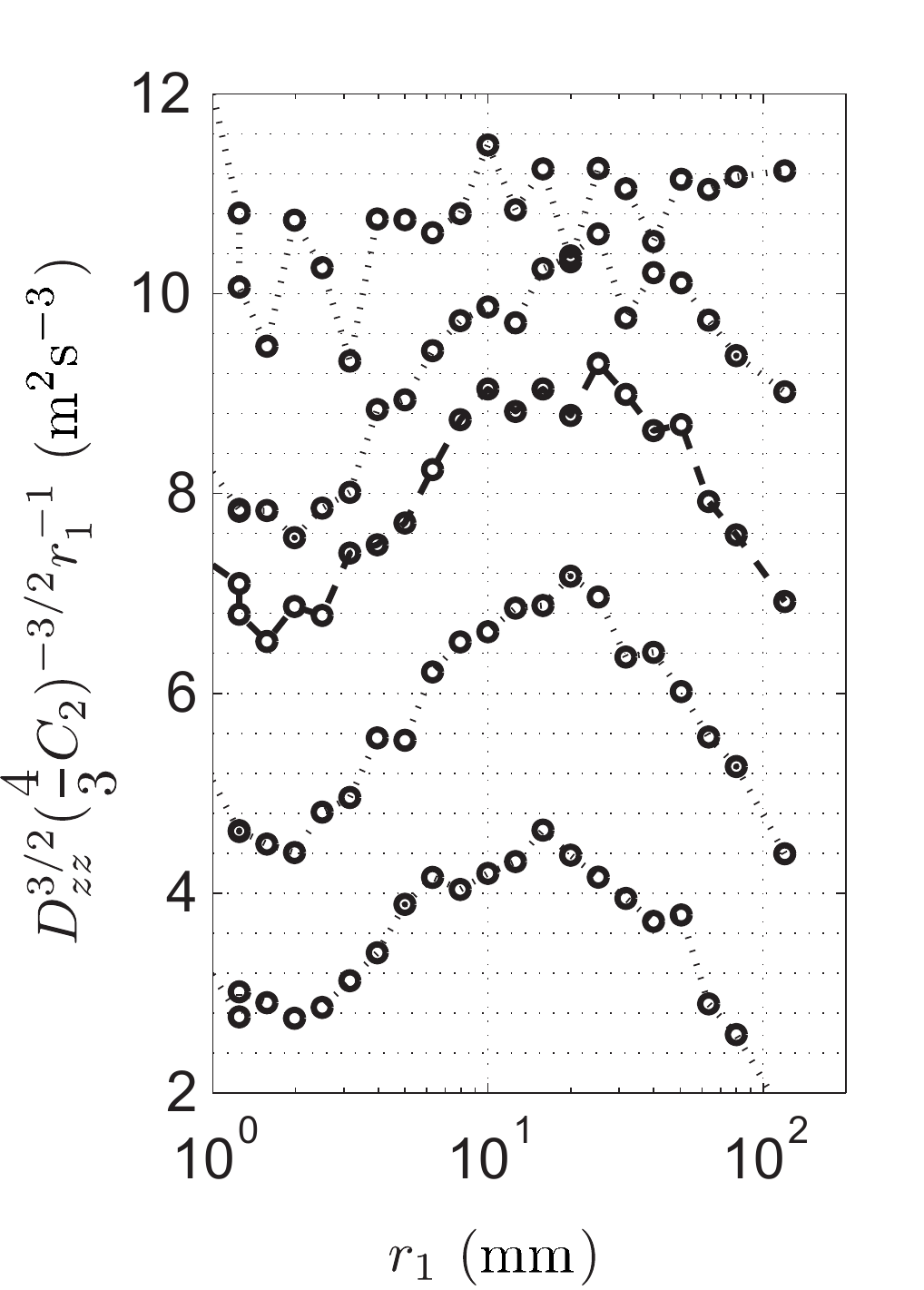}}
\caption{
 In (a) and (b), the axial structure functions, $D_{r_2 r_2} (z)$, 
 and radial structure functions, $D_{zz} (r_1)$, are normalized 
 by Kolmogorov's inertial range predictions given by equations~\ref{eq:drr} and \ref{eq:dzz}.  
 In order from bottom to top, the anisotropy, $\sigma_z/\sigma_r$, 
 was 0.59, 0.77, 0.94, 1.16 and 1.98.  
 Each curve was shifted by 1.5~m$^2$s$^{-3}$ with respect to the one below it 
 (except the bottom one).  
}
\label{fig:structure_func}
\end{center}
\end{figure}

We first use the transverse structure functions, 
defined by equations~\ref{eq:dzz} and \ref{eq:drr}, 
to estimate the Reynolds number of the turbulence.  
In figure~\ref{fig:structure_func}, the structure functions 
$D_{zz} (r_1)$ and $D_{r_2 r_2} (z)$ are normalized by the Kolmogorov scaling.  
That is, we solve the equations for $\epsilon$, the dissipation rate.  
Here, we assume that the Kolmogorov constant has the value given by \cite{sreenivasan:1995}, 
$C_2 = 2.1$, which is the mean of values taken from a collection of experimental studies.  
We assess the validity of this assumption later.  
We take the peak value of the function as our definition of the dissipation rate of the flow.  
The values estimated from $D_{r_2 r_2} (z)$ are given in table~\ref{table:velstat}.  
The corresponding values estimated from $D_{zz}(r_1)$ differ from those estimated from 
$D_{r_2 r_2} (z)$ by no more than 13\%.  
The dissipation rates depended slightly on the driving and reached a 
value of $6.7$~m$^{2}$s$^{-3}$ for the case of isotropic turbulence.  
Under isotropic forcing, the Taylor scale was then 
$\lambda = \sqrt{15 \, \nu \, \sigma^{2} / \epsilon} = 6.7$~mm, 
where $\nu$ is the kinematic viscosity of air at room temperature 
($1.57\times 10^{-5}$~m$^{2}$~s$^{-1}$), and 
the spherically averaged RMS velocity fluctuations 
was then $\sigma = \sqrt{(\sigma_z^2 + 2 \, \sigma_r^2)/3}$, 
which gave a Taylor-microscale Reynolds number of
$R_{\lambda} = \lambda \, \sigma / \nu = \reynoldsnum$.  
The Reynolds numbers for the anisotropic cases were similar.  

\begin{figure}
\begin{center}
\includegraphics[scale=0.75]{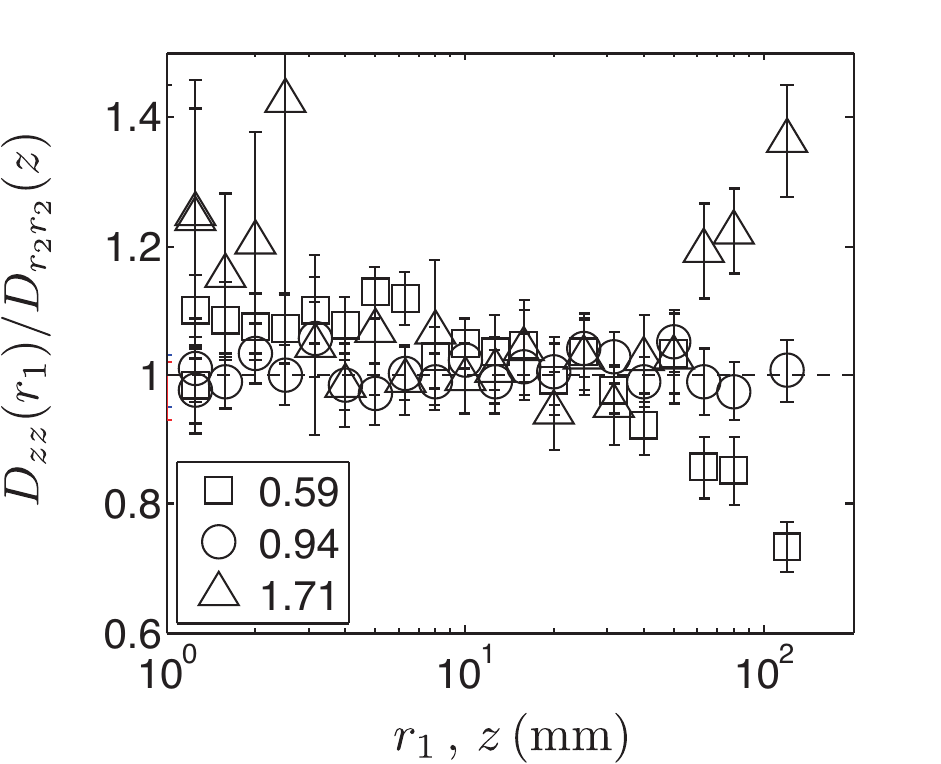}
\caption{
 The ratios between structure functions in different directions 
 but in the same flow, $D_{zz} (r_1) / D_{r_2 r_2} (z)$, 
 approach the isotropic value of one 
 as the separation $r_1 = z$ decreases.  
 Note that noise became important below separations of about 4~mm, 
 as can also be seen in figure~\ref{fig:structure_func}.  
 Results for data taken at other values of the large-scale anisotropy are 
 consistent with those shown here.  
 The values in the legends are the anisotropy, $\sigma_z/\sigma_r$, 
 measured at the centre of the soccer ball.  
}
\label{fig:structure_func_ratio}
\end{center}
\end{figure}

We defined a scale-dependent measure of anisotropy, 
$D_{zz} (r_1) / D_{r_2 r_2} (z)$ where $r_1 = z$.  
The data are shown in figure~\ref{fig:structure_func_ratio}.  
Local isotropy requires that 
$D_{zz} / D_{r_2 r_2}$ approaches one at small scales.  
In the limit of large separations, the ratio should approach 
$\sigma^2_z / \sigma^2_r$, since the velocities 
$u_i (\boldsymbol{r})$ and $u_i (0)$ 
are uncorrelated when $\boldsymbol{r}$ is large enough.  
Although we did not observe this regime, 
the values of the ratio in anisotropic cases did separate from the isotropic value 
as if to approach the values $\sigma^2_z / \sigma^2_r$ at larger, 
unobserved, scales.  
The figure shows that we observed local isotropy, within the experimental error, 
for separations smaller than about 50~mm.  
This approach to isotropy, then, is a consequence of the velocity correlations 
inherent in the dynamics of turbulence.  
We extract further information from the structure function ratio 
through two interpretations, described below.  

\begin{figure}
\begin{center}
\subfigure[]{
\includegraphics[scale=0.6]{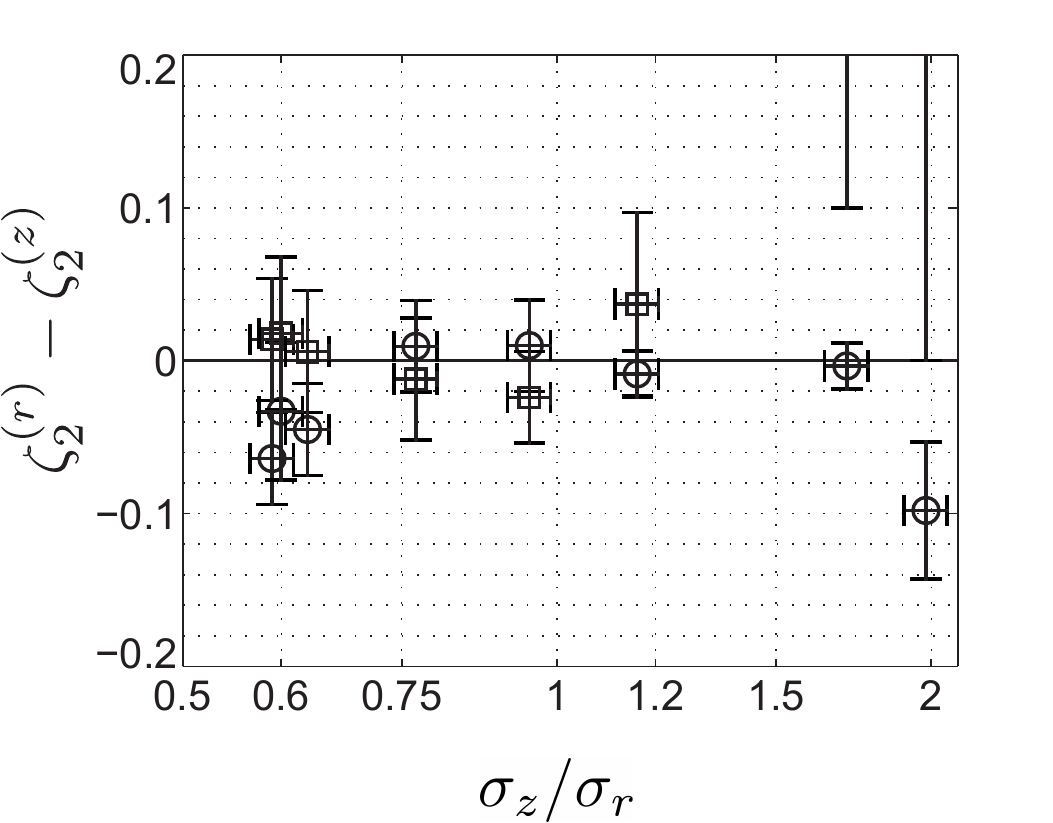}}
\hfill
\subfigure[]{
\includegraphics[scale=0.6]{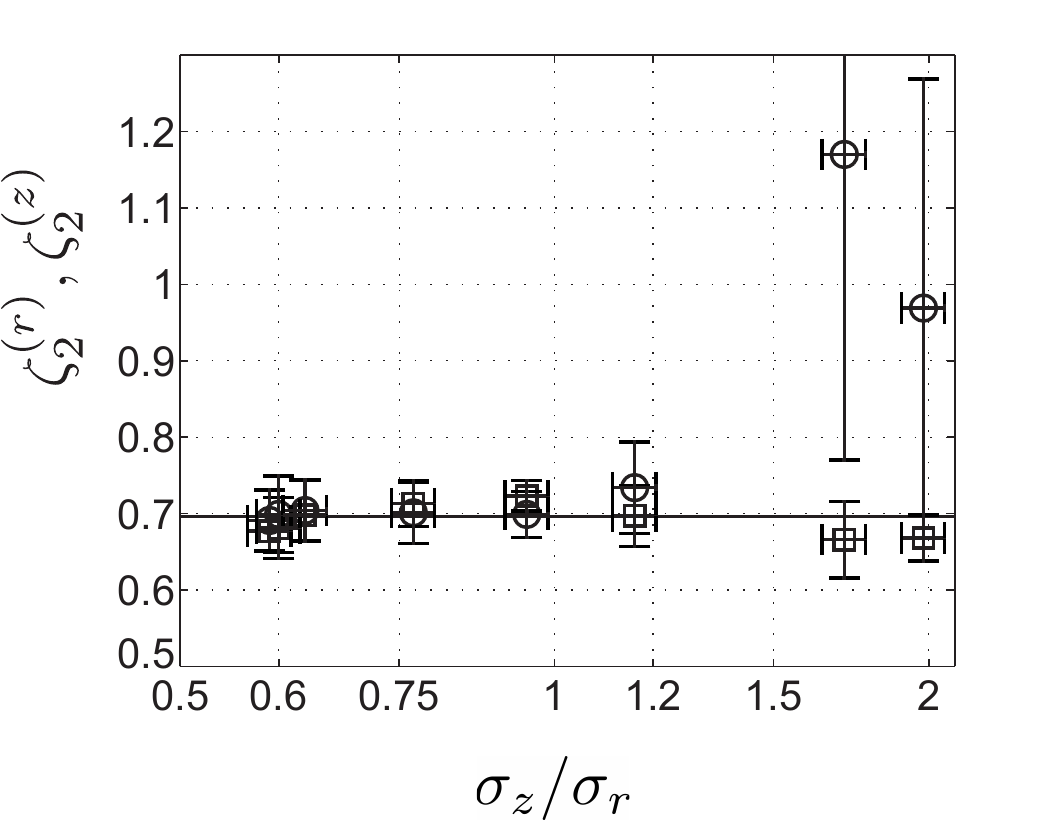}}
\caption{
 In (a), the ${\ocircle}$ symbols are the scaling exponent differences 
 measured through the use of equation~\ref{eq:Dratio}.  
 For the ${\square}$ symbols, the differences were computed 
 through the use of equation~\ref{eq:ESS}.  
 Two data points above the top of the plot and are out of view 
 in order to make the deviations of the other points more visible.  
 The values of these two points are 0.5$\pm$0.4 and 0.3$\pm$0.3 
 in order of increasing anisotropy.  
 The exponents measured using equation~\ref{eq:ESS} are shown in (b).  
 Here, the ${\ocircle}$ and ${\square}$ symbols mark radial and 
 axial scaling exponents, respectively.  
}
\label{fig:zeta}
\end{center}
\end{figure}

Because the structure functions do not exhibit clear scaling, 
as is evident in figure~\ref{fig:structure_func}, we employ ESS to 
extract scaling exponents.  
We employed it in a new way, to compare the structure functions 
measured in different directions, 
rather than to compare structure functions of different orders.  
According to \cite{benzi:1995}, ESS implies that 
\begin{equation}
D_{ii} = A^{(j)} \left(\frac{x_j}{L} \,\, f \left( \frac{x_j}{\eta} \right) \right)^{\zeta_2^{(j)}}, 
\label{eq:Dii}
\end{equation}
where $L$ is a large scale of the flow and $A^{(j)}$ are unknown constants.  
For standard ESS, one finds that the function $f(x_i/\eta)$ is independent 
of the order of the structure function.  
We assumed that it is also independent of the direction in which the structure 
function is measured.  
In this way, we assumed isotropy of the function $f$, but not of the underlying scaling.  
It follows that 
\begin{equation}
\frac{D_{zz} (r_1)}{D_{r_2 r_2} (z)} 
= \frac{ A^{(r)} \left( (r_1/L) \,\, f( r_1/\eta ) \right)^{\zeta_2^{(r)}} } 
{ A^{(z)} \left( (z/L) \,\, f( z/\eta ) \right)^{\zeta_2^{(z)}} } 
= A^\prime r_1^{\zeta_2^{(r)} - \zeta_2^{(z)} } 
= A^\prime r_1^{\Delta \zeta_2}.  
\label{eq:Dratio}
\end{equation}
whenever $r_1$ equals $z$.  
The ratio is a form of ESS because it eliminates the influence of the function $f$ 
in order to reveal the underlying scaling set by the exponents $\zeta_2^{(j)}$.  
To determine $\Delta \zeta_2$, 
we fit power laws to the data in figure~\ref{fig:structure_func_ratio} 
in the region between 4 and 40~mm, 
the lower bound being set by the appearance of noise in the data, 
and the upper bound by the emergence of the large-scale cutoff.  
It can be seen in figure~\ref{fig:zeta}(a) 
that the dependence of $\Delta \zeta_2$ on the anisotropy 
is non-monotonic.  

\begin{figure}
\begin{center}
\subfigure[]{
\includegraphics[scale=0.75]{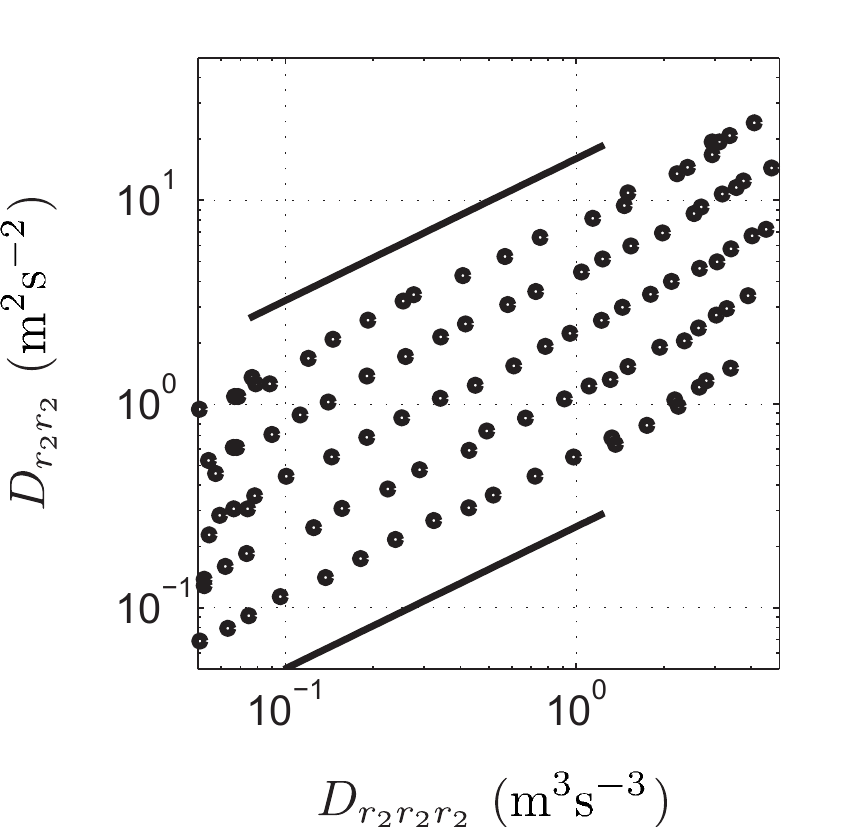}}
\hfill
\subfigure[]{
\includegraphics[scale=0.75]{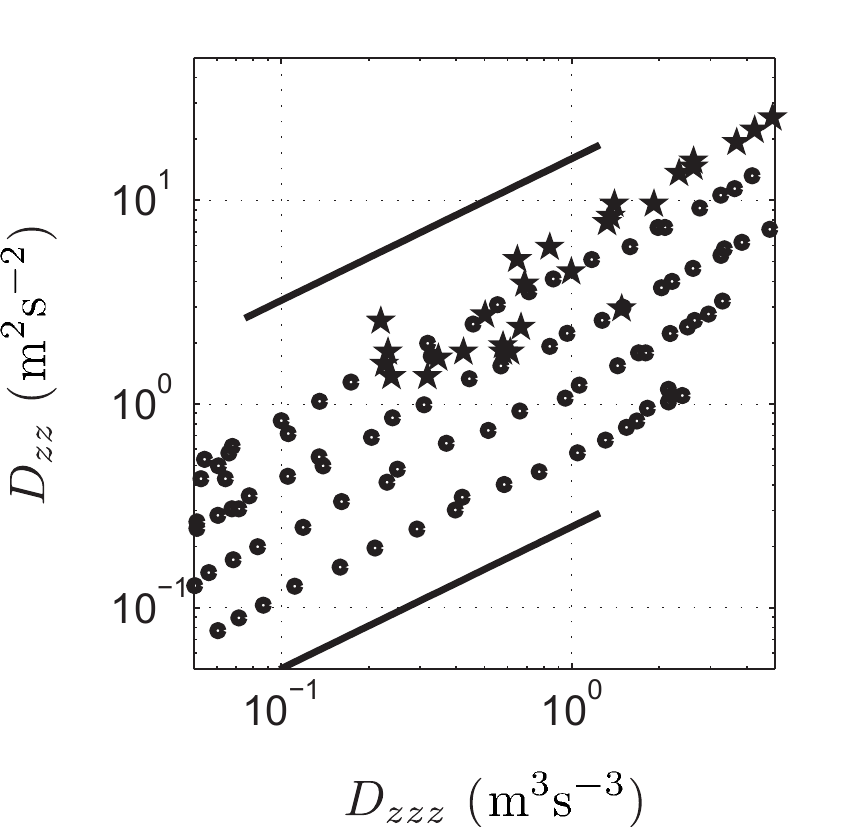}}
\caption{
 In (a) and (b), the axial structure functions, $D_{r_2 r_2} (z)$, 
 and radial structure functions, $D_{zz} (r_1)$, are plotted against 
 their third-order counterparts.    
 For each curve in order from bottom to top, the anisotropy, 
 $\sigma_z/\sigma_r$, was 0.59, 0.77, 0.94, 1.16 and 1.98.  
 Each curve was shifted by a factor of 2 with respect to the one below it 
 (except the bottom one).  
 The solid lines are power laws with an exponent of 0.70.  
 The uppermost curve in (b) is marked with stars in order to make the data 
 distinguishable from the other data.  
 For these data, the third order structure function did not converge.  
}
\label{fig:ess}
\end{center}
\end{figure}

Figure~\ref{fig:zeta}(a) also shows exponent differences derived from standard ESS.  
In the original expression of ESS, 
\begin{equation}
D_{ii} \sim D_{iii}^{\zeta_2^{(j)}}, 
\label{eq:ESS}
\end{equation}
where $D_{iii}$ is the third moment of the absolute value of the velocity difference, 
$\langle |u_i (x_j) - u_i (0)|^3  \rangle$.  
For ESS to be present, $D_{iii} = B_i \, (r/L) \, f(r/\eta)$.  
In figure~\ref{fig:ess}, we plot $D_{zz}$ against $D_{zzz}$, and $D_{r_2 r_2}$ 
against $D_{r_2 r_2 r_2}$, in order to uncover the scaling exponents 
$\zeta_2^{(r)}$ and $\zeta_2^{(z)}$, respectively.  
We fit power laws to the data in figure~\ref{fig:ess} 
for $r$ and $z$ between 4 and 40~mm, as before.  
The exponents of these power laws are shown in figure~\ref{fig:zeta}(b).  
Note that for two exponents measured in the radial direction of prolate spheroidal turbulence, 
the data were insufficient to bring about convergence of the third order structure functions.  
This is evident in the scatter present in the corresponding data in figure~\ref{fig:ess}.  
We note speculatively that for these two conditions, the Reynolds stresses were about 0.08 
higher than for all of the other cases, as seen in table~\ref{table:velstat}, 
and that this may point to a role for shear in altering the scaling.  

As seen in figure~\ref{fig:zeta}(b), 
the dependence of the scaling exponents,  $\zeta_2^{(r)}$ and $\zeta_2^{(z)}$, 
on anisotropy was non-monotonic.  
Furthermore, deviations of $\Delta \zeta_2$ from zero measured using standard ESS 
do not follow the same trend as the those measured through equation~\ref{eq:Dratio}.  
This discrepancy suggests that anisotropy is by itself not responsible for the deviations, 
but that they have their origin in measurement noise and error.  
The mean value of the scaling exponents, excluding the two outliers, was 0.70$\pm$0.03.  
This value is the same as the one found by \cite{benzi:1995} in a wind tunnel at 
comparable Reynolds number, 
and the value is indicated with a horizontal line in the figure.  

\cite{shen:2002} found that the exponent of the second order transverse structure function 
was 0.1 smaller in the direction normal to the mean flow in a wind tunnel, both with 
and without shear.  
Wind tunnel turbulence is approximately axisymmetric in the absence of shear, 
with the axis of the tunnel corresponding to our $z$.  
Although one of our data points, namely the one for 
$\sigma_z/\sigma_r$~=~1.98, does fall 
at $-0.1$, in agreement with \cite{shen:2002}, their result cannot be reconciled with 
the body of our data.  

Let us consider the Kolmogorov constant.  
We calculated the ratio $C_{2}^{(r)} / C_{2}^{(z)}$ in two ways.  
First, we computed an average of the structure function ratio, 
$\langle D_{zz} (r) / D_{r_2 r_2} (z) \rangle$, shown in figure~\ref{fig:structure_func_ratio}.  
The average was taken over a range of separations that 
bracketed the peak in the compensated structure functions, 
or $4 {\rm mm} < r_1, z < 40 {\rm mm}$.  
In the inertial range, according to equations~\ref{eq:dzz} and \ref{eq:drr}, 
\begin{equation}
\frac{ D_{zz}(r_1) }{ D_{r_2 r_2} (z) } 
= \frac{ \tfrac{4}{3} C_2^{(r)} (\epsilon r_1)^{\zeta_2^{(r)}} }{ \tfrac{4}{3} C_2^{(z)} (\epsilon z)^{\zeta_2^{(z)}} }
= \frac{ C_2^{(r)} }{ C_2^{(z)} }
\label{eq:Cratio}
\end{equation}
when $r_1 = z$, $\zeta_2^{(r)} = \zeta_2^{(z)}$,   
and because the dissipation, $\epsilon$, is a scalar quantity.  
We used the standard deviation in the value of the ratio 
as a measure of the error of the measurement.  
The second method was designed to mimic the one usually 
employed to measure $C_2$ (or $\epsilon$) 
when data are collected in only one direction.  
That is, we estimated each $C_2$ from the maximum of the compensated structure functions.  
In order to reduce the influence of noise, we found the maximum of a polynomial function, 
second-order in $\log (r)$, fit to the data between 4 and 40~mm.  
The ratio between the maxima, 
$\max (D_{zz} (r) / r^{2/3}) / \max (D_{r_2 r_2} (z) / z^{2/3})$, 
was then calculated.  
As before, the quantity is an estimate for the ratio of Kolmogorov constants 
because the dissipation rate cancels.  
We used the deviation of the data from the fitted parabolas to estimate the error.  
Although each method described here suffers from limitations, 
they both serve as practical definitions of $C_{2}^{(r)} / C_{2}^{(z)}$.  
\begin{figure}
\begin{center}
\includegraphics[scale=0.65]{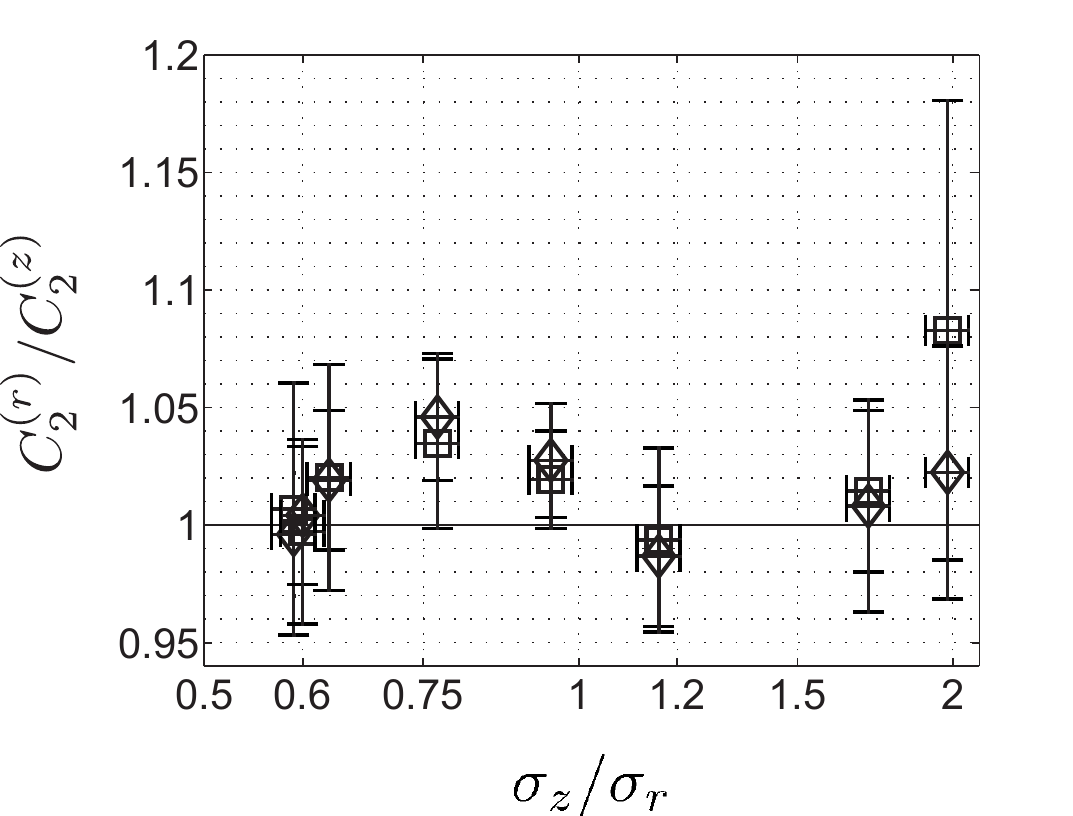}
\caption{
The ratio between Kolmogorov constants measured in different directions, 
$C_{2}^{(r)} / C_{2}^{(z)}$.  
The ${\Diamond}$ symbols are the ratio between the maximum values 
of the compensated radial and axial structure functions.  
The ${\square}$ symbols are $\langle D_{zz} (r_1) / D_{r_2 r_2} (z) \rangle$, 
the ratio of radial to axial structure functions averaged 
over a range of scales $4 < r_1, z < 40~{\rm mm}$.  
}
\label{fig:c2ratio}
\end{center}
\end{figure}

The two measures of $C_{2}^{(r)} / C_{2}^{(z)}$ are plotted in figure~\ref{fig:c2ratio}.  
The ratios deviated from the isotropic value by up to 4\%, except for one outlier 
whose value was about 8\% higher than 1.0.  
The observed dependence on the anisotropy, if there was one, was non-monotonic.  
However, \cite{busa:1997} found in a shell model of turbulence 
that the Kolmogorov constant varied monotonically with anisotropy.  
Although they predicted a dependence too weak to detect in our data, 
it is reasonable to expect one that is qualitatively similar.  
Since we did not see such a monotonic dependence, 
we concluded that the deviations of the Kolmogorov constants from isotropy 
were probably not due to the anisotropy in the fluctuations, 
but were rather the result of measurement noise and error.  
Though large, the uncertainty in our data is smaller than the scatter 
seen in the data collected by \cite{sreenivasan:1995}, whose amplitude is about 10\%.  
It follows that anisotropy in the velocity fluctuations alone does not account 
for the variation in previously measured values of the Kolmogorov constant.

\section{Conclusions}
\label{sec:conclusions}

We investigated systematically the influence of anisotropic agitation 
on the inertial scales of turbulence in an experiment with Taylor-based 
Reynolds number $R_\lambda = \reynoldsnum$.  
Thirty two loudspeaker-driven jets pointed toward the centre of a spherical chamber  
and were driven to produce axisymmetric turbulence, 
which in a central volume had no shear, no mean flow, and was homogeneous.  
We observed flows for which the ratio of axial to radial RMS velocity fluctuations 
was between 0.6 and 2.3.  
We found that the anisotropy of the velocity fluctuations at the largest scales 
had the same anisotropy as the agitation.  
According to two inertial range measures, the second order velocity structure 
functions were independent of anisotropy.  
There was extended self-imilarity, 
and the structure functions had the same scaling exponent 
in different directions and for different anisotropies, 
0.70$\pm$0.03.  
The Kolmogorov constant, $C_2$, also showed no dependence on the anisotropy.  
Because of this, we expect anisotropic corrections, such as those uncovered by 
the SO(3) decomposition \cite[e.g.][]{kurien:2001, biferale:2005}, 
to be smaller than the error in our measurements (about 4\%).  
This holds unless the anisotropic corrections cancelled 
in the two directions we measured, an outcome we consider unlikely.  
Further experiments are necessary to determine whether deviations   
from isotropy present within the uncertainty of our measurement are significant.  

\begin{acknowledgments}
This research was supported by the Max Planck Society and was carried 
out in cooperation with the International Collaboration for Turbulence Research.  
We thank Lance Collins, Zellman Warhaft, K.R. Sreenivasan and Reginald Hill 
for valuable suggestions.  
We thank Mathieu Gibert for drawing our attention to the work of \cite{fox:1988}.  
Kelken Chang acknowledges financial support from 
the Deutsche Forschungsgemeinschaft (German Science Foundation) 
through the grant XU91/3-1.  
\end{acknowledgments}

\bibliographystyle{jfm}
\bibliography{bibliography}

\end{document}